\documentclass[fleqn,usenatbib]{mnras}

\usepackage{newtxtext,newtxmath}

\usepackage[T1]{fontenc}

\DeclareRobustCommand{\VAN}[3]{#2}
\let\VANthebibliography\thebibliography
\def\thebibliography{\DeclareRobustCommand{\VAN}[3]{##3}\VANthebibliography}

\usepackage{graphicx}	
\usepackage{amsmath}	

\title[Ambipolar, Hall and Ohmic evolution]{Combined magnetic field evolution in neutron star cores and crusts: Ambipolar diffusion, Hall effect and Ohmic dissipation}

\author[D. Skiathas and K. N. Gourgouliatos]{
Dimitrios Skiathas,$^{1,2,3}$\thanks{E-mail: d.skiathas@ac.upatras.gr, dimitrios.skiathas@nasa.gov}
Konstantinos N. Gourgouliatos,$^{1}$\thanks{E-mail: kngourg@upatras.gr}
\\
$^{1}$ Department of Physics, University of Patras, Patras, Rio, 26504, Greece\\
$^{2}$ Southeastern Universities Research Association, Washington, DC 20005, USA\\
$^{3}$ Astrophysics Science Division, NASA/Goddard Space Flight Center, Greenbelt, MD 20771, USA
}

\date{Accepted 2024 January 11. Received 2023 December 6; in original form 2023 September 1}

\pubyear{2024}

\begin{document}
\label{firstpage}
\pagerange{\pageref{firstpage}--\pageref{lastpage}}
\maketitle

\begin{abstract}
Neutron star magnetic field evolution is mediated through the Hall effect and Ohmic dissipation in the crust while ambipolar diffusion is taking place in the core. These effects have been studied in detail in either part of the star, however, their combined, simultaneous evolution and interplay has not been explored in detail yet. Here, we present simulation results of the simultaneous evolution of the magnetic field in the core due to ambipolar diffusion and the crust due to Hall effect and Ohmic decay, under the assumption of axial symmetry. We find that a purely poloidal field generates a toroidal field in the crust, due to the Hall effect, that sinks into the core. A purely toroidal field remains toroidal and spreads into the core and the crust. Finally, for a mixed poloidal-toroidal field, the north-south symmetry is broken due to the Hall effect in the crust, however, ambipolar diffusion, tends to restore it. We examine the role of ambipolar diffusion to the magnetic field decay and we compare the rate of the conversion of magnetic field energy into heat, finding that it enhances the magnetic field decay in neutron stars. 

\end{abstract}




\section{Introduction}

Neutron stars manifest a wide variety of observable behaviours that leads to categorisation in different classes. This observed diversity poses the question of the main cause of these differences. The progress towards their Grand Unification \citep{Kaspi2010} shows that the main reason behind this diversity is their magnetic field \citep{Harding2013}. The most extreme incarnation of neutron stars, magnetars, are highly magnetized ones with a large reservoir of energy that is released in the form of burst and flares. The rapid onset of these events and the long term persistent thermal emission, deem necessary that their magnetic field evolves, magnetic energy is converted into heat triggering these explosive events. High magnetic field pulsars, occasionally exhibit magnetar-like behaviour, and transition between magnetar and pulsar states \citep{2008Sci...319.1802G,2010Sci...330..944R,2016ApJ...829L..21A}. Even in the weak-end dipole magnetic field, central compact objects are bright in X-rays, yet their timing properties hint towards relatively slow rotational velocities and mild spin-down rate \citep{1980ApJ...239L.107T, 2006Sci...313..814D, 2015PASA...32...18P,2019A&A...626A..19E}. In this context, a global and unified picture of the main mechanisms behind magnetic field evolution of neutron stars is essential. 

\cite{Goldreich1992} set the main framework for studies related to neutron stars magnetic field evolution. The Hall effect and subdominant Ohmic dissipation are responsible for the evolution in the crust while ambipolar diffusion is the dominant mechanism in the core of highly magnetized ones, i.e.~magnetars with field strengths of $10^{14}-10^{15}$ G. In cases with appropriate conditions in the core, effects due to superconductivity and superfluidity also contribute to the evolution of the magnetic field \citep{Glampedakis2011,2013PhRvL.110g1101L,2014MNRAS.437..424L,Graber2015,Passamonti2017,Dommes2017}. 

The evolution in the crust under the Hall effect and Ohmic dissipation has been studied in detail both analytically and numerically \citep{Hollerbach2002, Hollerbach2004,2004ApJ...609..999C,  Reisenegger2007etal, Pons2009,  Vigano2013}. In axial symmetry the field is found to evolve toward the Hall attractor state \citep{Gourgouliatos2014b,Gourgouliatos2013}. In three dimensions  \cite{Gourgouliatos2016} found that non-axisymmetric features can developed even from an initially axisymmetric configuration. Moreover, more complex states involving spots, zones of magnetic field and turbulent fields  have been studied in this context \citep{2014MNRAS.444.3198G,2015PhRvL.114s1101W, Geppert2017, Gourgouliatos2018, Pons2019, Gourgouliatos2020, Igoshev2021b, Dehman2023}  and more recently the case of off-centered dipolar and multipolar magnetic fields and their implications on neutron star evolution has been explored \citep{Igoshev2023,2023MNRAS.522.5879A}.

The mechanism of ambipolar diffusion in the core of a neutron star has also be studied, yet, its non-linear dependence on the magnetic field through the cubic power, and the complication of microphysics of the inner crust and the core, have made this task rather challenging. There are studies that have examined the impact of ambipolar diffusion in a core of a neutron star consisting of normal matter, in one dimension \citep{Hoyos2008}, in axial symmetry \citep{Castillo2017,Castillo2020,Passamonti} and more recently in three dimensions \citep{Igoshev2022}.  The effect of superfluidity and superconductivity has also been examined both in some of the works already referred but also in the works of \cite{Elfritz2016,Glampedakis2011b,Bransgrove2017}. However the relevance and the impact these effects have, in magnetic field evolution, is still not completely resolved. 

The prevailing approach of these studies is that the evolution is simulated in detail in the region under consideration, i.e.~the crust or the core. However, the state of the magnetic field in the rest of the star, is approximated through either a stationary magnetic field, or, in some cases is devoid of magnetic field, using an appropriate assumption, i.e.~the Meissner effect has pushed the field out of the core. 
There are few studies considering the coupled evolution of the crust under Hall effect and Ohmic dissipation and the core magnetic field \citep{Bransgrove2017,Elfritz2016,Vigano2021}. In the first two the core magnetic field evolution is mainly driven by the interaction of the flux tubes with the core matter or the neutron vortices; in the third one the primary focus is on the magnetothermal evolution with the inclusion of the mechanism of ambipolar diffusion, for relatively short timescales.

As the region of the crust-core interface containts a highly resistive layer \citep{2013NatPh...9..431P}, the formation of strong currents there leads to efficient dissipation or the generation of instabilities \citep{2002PhRvL..88j1103R,2004A&A...420..631R,2010A&A...513L..12P,2014PhPl...21e2110W,2016MNRAS.463.3381G}, which however require fully three-dimensional calculations for their study \citep{Gourgouliatos2020}. Simulations that focus exclusively either in the crust or the core cannot demonstrate the full range of these phenomena, nevertheless, such effects are, in some occasions introduced, in the context of bursts and glitches \citep{2020ApJ...897..173B,2021ApJ...908..176Y}.  

In this work we present the results of a novel code that simulates the evolution of the magnetic field in a highly magnetized neutron star. We take into account both the crust, where the magnetic field evolves due to the Hall effect and Ohmic dissipation, and the core where ambipolar diffusion dominates. These effects are simulated simultaneously. We do not account for the effects of superconductivity and superfluidity. The physical parameters determining the pace of the evolution (Hall and ambipolar diffusion parameter) are piece-wise constant in the crust and the core and invariable with respect to time. 

The structure of this paper is as follows. In section 2 we describe the physical model, extracting the evolution equation and the relevant conditions that need to be satisfied.  The cases studied and the corresponding results of the simulations are presented in section 3. In section 4 we discuss the results comparing them with previous studies. In section 5 we summarise our conclusions.

\section{Method}

\subsection{Equations of motion}
\label{sec:maths} 
We are considering a neutrally charged non-rotating neutron star consisting of normal matter. The interior of the neutron star is modeled as a mix of protons, neutrons and electrons, labeled by the indices $ \mathrm n, \mathrm p, \mathrm e$. For example the number density of the involved particles would be $n_{\mathrm n}, n_{\mathrm p}$ and $n_{\mathrm e}$ respectively.
The local charge neutrality of the neutron star implies that $n_{\mathrm e}=n_{\mathrm p}$. Hereafter $n_{\mathrm c}$ would be the number density of charged particles. The total number of baryons \textbf{is} $n_{\mathrm b}$ and the fraction of neutrons to baryons is $x_{\mathrm n}$. In our calculations and simulations the neutron number density is considered to be uniform in the entire star.

Charged particles interact through electromagnetic forces. Between electrons and neutrons there are weak interactions, while protons and neutrons are subject to nuclear forces. All particles are under the influence of a gravitational potential. 

Based on the model described by \citet{Goldreich1992} the equation of motion for the electrons, protons and neutrons are:
\begin{equation}
\begin{split}
    m_e^*\frac{\partial\bmath{\upsilon_e}}{\partial t}+m_e^*(\bmath{\upsilon_e}\cdot\bmath{\nabla})\bmath{\upsilon_e} =&-\bmath{\nabla}\mu_e-m_e^*\bmath{\nabla}\Phi-e(\bmath{E}+\frac{\bmath{\upsilon_e}}{c}\times\bmath{B})\\ 
    &-\frac{m_e^*(\bmath{\upsilon_e}-\bmath{\upsilon_n})}{\tau_{en}}-\frac{m_e^*(\bmath{\upsilon_e}-\bmath{\upsilon_p})}{\tau_{ep}}
    \end{split}
    \label{eqm1}
\end{equation}
\begin{equation}
\begin{split}
    m_p^*\frac{\partial\bmath{\upsilon_p}}{\partial t}+m_p^*(\bmath{\upsilon_p}\cdot\bmath{\nabla})\bmath{\upsilon_p} =&-\bmath{\nabla}\mu_p-m_p^*\bmath{\nabla}\Phi+e(\bmath{E}+\frac{\bmath{\upsilon_p}}{c}\times\bmath{B})\\ &-\frac{m_p^*(\bmath{\upsilon_p}-\bmath{\upsilon_n})}{\tau_{pn}}-\frac{m_p^*(\bmath{\upsilon_p}-\bmath{\upsilon_e})}{\tau_{pe}}
     \end{split}
     \label{eqm2}
\end{equation}
\begin{equation}
\begin{split}
    m_n^*\frac{\partial\bmath{\upsilon_n}}{\partial t}+m_n^*(\bmath{\upsilon_n}\cdot\bmath{\nabla})\bmath{\upsilon_n} 
    =&-\bmath{\nabla}\mu_n-m_n^*\bmath{\nabla}\phi \\
    &-\frac{m_n^*(\bmath{\upsilon_n}-\bmath{\upsilon_p})}{\tau_{np}}-\frac{m_n^*(\bmath{\upsilon_n}-\bmath{\upsilon_e})}{\tau_{ne}}
    \end{split}
    \label{eqm3}
\end{equation}
where $\bmath{E}$ is the electric field, $\bmath{B}$ is the magnetic field, $\Phi$ is the gravitational potential, $\tau_{\alpha\beta}$ is the relaxation time for collision between $\alpha$ and $\beta$ particles (by $\alpha$ and $\beta$ we denote any species of particles involved in some interaction), $c$ is the speed of light, $e$ the electron charge, $\bmath{ \upsilon_\alpha}$ is the velocity of $\alpha$ particles and  $m_\alpha^*$ is the effective mass of $\alpha-$particles.

The electron effective mass is defined as $m^*_{\mathrm e}=m_{\mathrm e}(1+x_e^2)^{1/2}$, where $x_{\mathrm e}$ is the ratio of the Fermi momentum to electron rest mass. On the other hand because protons are not degenerate relativistic particles in the core of neutron stars like electrons, their effective mass is smaller than their rest mass. We take protons effective mass to be constant and equal to $0.6m_{\mathrm p}$ where $m_{\mathrm p}$ is the proton rest mass. In a similar way the neutron effective mass is defined. In the calculation we perform here, the conservation of momentum is used where $n_\alpha m^*_\alpha/\tau_{\alpha\beta}=n_\beta m^*_\beta/\tau_{\beta\alpha}$  and eventually only the proton effective mass will be used in the final expressions.

The acceleration terms in the left side of equations (\ref{eqm1}-\ref{eqm3}) can be neglected in the framework of quasi-stationary evolution driven by slow motions.  Under the hypothesis of stationary neutrons their friction with the charged particles provide the necessary mechanism for the charged particles to evolve through successive quasi-equilibrium states without the need to introduce artificial frictional forces to restore the hydro-magnetic quasi-equilibrium as used in \citet{Hoyos2008} and \citet{Castillo2020}, where moving neutrons are considered.

\subsection{Induction Equation}

Using the method described in \citet{Goldreich1992} and \cite{Passamonti} we can combine equations (\ref{eqm1}),(\ref{eqm2}) and (\ref{eqm3}) to obtain Ohm's law for the electric field. Then using Faraday's law $\partial\bmath{B}/\partial t=-c\bmath{\nabla}\times\bmath{E}$ we take the evolution equation for magnetic field
\begin{equation}
    \frac{\partial \bmath{B}}{\partial t} = -\bmath{\nabla}\times\Big(\frac{c}{4\pi en_c}(\bmath{\nabla}\times\bmath{B})\times\bmath{B}+\frac{c^2}{4\pi\sigma}\bmath{\nabla}\times\bmath{B}\Big)+\bmath{\nabla}\times(\bmath{\upsilon_p}\times\bmath{B})
    \label{eqb}
\end{equation}
The above equation is the induction equation where the first two terms describe the evolution of magnetic field due to Hall effect and Ohmic dissipation respectively. 

\subsection{Ambipolar diffusion velocity}
\label{Sec:Amb_Vel}

In the last term of equation (\ref{eqb}),  $\mathbf{\upsilon_p}$ is not the ambipolar diffusion velocity but the velocity of protons. As explained in \citet{Passamonti} and \citet{Igoshev2022} we have to expand $\bmath{\upsilon_p}$ as $\bmath{\upsilon_p}=\bmath{\upsilon_b} + x_n(\bmath{\upsilon_p}-\bmath{\upsilon_n})$ where $\bmath{\upsilon_b}$ is the baryon velocity 
defined as $\bmath{\upsilon_b}=(n_n\bmath{\upsilon_n}+n_p\bmath{\upsilon_p})/n_b$. Although recent studies found baryon velocity to be large \citep{Castillo2020,2018PhRvD..98d3007O}, we assume here that this velocity is negligible similar to \citep{Castillo2017,Passamonti,Igoshev2022}, thus $\bmath{\upsilon_p}=\bmath{\upsilon_{amb}}$ where we have defined ambipolar diffusion velocity to be $\bmath{\upsilon_{amb}}=x_n(\bmath{\upsilon_p}-\bmath{\upsilon_n})$. This assumption although it is not changing qualitatively the evolution, it makes it slower, resulting in longer timescales.

Combining  equations (\ref{eqm1}), (\ref{eqm2}) and (\ref{eqm3}) as shown in \citet{Passamonti} and analytically explained by \citet{Igoshev2022} in their Appendix A, the ambipolar diffusion velocity can be calculated by the following equation
\begin{equation}
    \bmath{\upsilon_{amb}}=\frac{x_n^2t_{pn}}{m_p^*}\Big[\frac{\bmath{f_{mag}}}{n_c}-\bmath{\nabla}(\Delta\mu)\Big]
    \label{vequation}
\end{equation}
The first term in the right-hand side is the magnetic force $\bmath{f_{mag}}={\bmath{j}\times\bmath{B}}/{c}$ where $\bmath{j}=c\bmath{\nabla}\times\bmath{B}/4\pi$ is the electric current. The second term is the deviation from chemical equilibrium $\Delta\mu=\mu_p+\mu_e-\mu_n$. 

In the core of the neutron star, weak interactions act to restore the chemical equilibrium. However, weak interactions will be fast enough only in the early life of a neutron star. In this case the departures from chemical equilibrium do not last long and  the gradient in (\ref{vequation}) is sufficiently small to be ignored. As a result ambipolar diffusion will be proportional to the magnetic force. 

Later as the neutron star cools down, weak interactions will not be sufficiently fast to restore the chemical equilibrium in short timescales and thus the magnetic force, see equation (\ref{vequation}), will be balanced by the gradient of the departures from chemical equilibrium.  These departures from chemical equilibrium can be calculated by solving the equation
\begin{equation}
\nabla^2(\Delta\mu)-\frac{m_p^*\lambda}{x_n^2 n_c t_{pn}} \Delta\mu=\bmath{\nabla}\cdot\left(\frac{\bmath{f_{mag}}}{n_c}\right)
\label{eqdm}
\end{equation}
where $\lambda$ is the change of reaction rate depending on deviation from chemical equilibrium $\lambda=(d\Gamma/\Delta\mu)|_{eq}$.
The derivation of equation (\ref{eqdm}) can be found in the Appendixes of \cite{Passamonti} or \cite{Igoshev2022}. The only difference here is that $n_c$ and the other physical parameters are considered uniform, thus the terms involving their spatial derivatives are zero.  

Considering equation (\ref{vequation}) we have to note that only the irrotational part of the magnetic force can be balanced, as the gradient of a scalar function is involved. Thus, departures from chemical equilibrium that may develop during the evolution of the system, will suppress the irrotational part of $\bmath{\upsilon_{amb}}$, thus, the velocity will be predominantly solenoidal \citep{Castillo2017,Passamonti}. 

In our simulations we choose the rate of weak interactions $\lambda$ to correspond to the one of the modified Urca reactions which given by \cite{Sawyer1989} as
\begin{equation}
    \lambda=5\times10^{27}T_8^6\Big(\frac{\rho}{\rho_{0}}\Big)^{2/3}erg^{-1}cm^{-3}s^{-1}
\end{equation}
where $\rho_0$ is the nuclear saturation density, $\rho$ neutron star density and $T_8$ the temperature in units of $10^8$K.

\subsection{Dimensionless Equations}

Using equation (\ref{vequation}) for ambipolar velocity and substituting the magnetic force and the electric current, the evolution equation (\ref{eqb}) for the magnetic field takes the form 
\begin{equation}
\begin{split}
    \frac{\partial \bmath{B}}{\partial t} = & -\bmath{\nabla}\times\Big[\frac{c}{4\pi en_c}(\bmath{\nabla}\times\bmath{B})\times\bmath{B}+\frac{c^2}{4\pi\sigma_0}\bmath{\nabla}\times\bmath{B}\Big] \\
    &+\bmath{\nabla}\times\Bigg[\frac{x_n^2t_{pn}}{m_p^*}\Big[\frac{(\bmath{\nabla}\times\bmath{B})\times\bmath{B}}{4\pi n_c}-\bmath{\nabla}(\Delta\mu)\Big]\times\bmath{B}\Bigg]
    \end{split}
    \label{Bequation}
\end{equation}
Setting $\bmath{b}=\bmath{B}/B_0$, $r=r/R_0$, $\tau=t/\tau_{B}$ with $\tau_{B}= 4\pi m_p^* n_c R_0^2/(x_n^2 t_{pn} B_0^2) $ and $Y=(4\pi n_c/B_0^2)\Delta\mu$ equation (\ref{Bequation}) becomes
\begin{equation}
\begin{split}
\frac{\partial\bmath{b}}{\partial\tau}=&-D_n\bmath{\nabla}\times(\bmath{\nabla}\times\bmath{b})-G_n\bmath{\nabla}\times[(\bmath{\nabla}\times\bmath{b})\times\bmath{b}] \\
&+\bmath{\nabla}\times\big[[(\bmath{\nabla}\times\bmath{b})\times\bmath{b}-\bmath{\nabla}Y]\times\bmath{b}\big]
\label{bfeq}
\end{split}
\end{equation}
where 
\begin{equation}
    G_n=\frac{cm_p^*}{x_n^2t_{pn}B_0e}
\end{equation}
and
\begin{equation}
    D_n=\frac{c^2m_p^*n_c}{\sigma x_n^2 t_{pn} B_0^2}
\end{equation}
In a similar way, an equation for $Y$ can be deduced starting from the equation of the deviations from chemical equilibrium (\ref{eqdm}). Using the dimensionless variables $b$ and $r$ and substituting the magnetic force we arrive at the following equation for $Y$
\begin{equation}
\nabla^2Y-A_cY=\bmath{\nabla}\cdot [(\bmath{\nabla}\times\bmath{b})\times\bmath{b}]
\label{Yequation}
\end{equation}
where 
\begin{equation}
    A_c=\frac{ R_0^2\lambda m_p^*}{n_c x_n^2t_{pn}}
\end{equation}

\subsection{Magnetic Field Configuration}

We restrict ourselves to axial symmetry, thus we can decompose the magnetic field in a poloidal and a toroidal component
\begin{equation}
    \bmath{B}=\bmath{\nabla}P\times\bmath{\nabla}\phi+T\bmath{\nabla}\phi
    \label{decompotion}
\end{equation}
The scalar functions $P(t,r,\theta)$ and $T(t,r,\theta)$ are associated to the poloidal and toroidal magnetic field respectively, $t$ denotes time, $r$ is the radial coordinate, $\theta$ the polar angle and $\phi$ the azimuthial angle. In spherical coordinates $\bmath{\nabla}\phi=(\hat{\bmath{\phi}}/r\sin{\theta})$.

\subsection{Final equations}
To derive the equations of evolution of the magnetic scalar functions we notice that the right-hand side of equation (\ref{bfeq}) is the curl of a vector quantity $\bmath{A}$, with 
\begin{equation} 
\bmath{A}=\bmath{A}^{Ohmic} +\bmath{A}^{Hall}+\bmath{A}^{AD}
\label{aeq}
\end {equation}
where
\begin{equation}
\bmath{A}^{Ohmic}= -D_n\bmath{\nabla}\times\bmath{b}
\end{equation}
is a vector quantity, whose curl describes the evolution of the magnetic field under Ohmic dissipation
\begin{equation}
    \bmath{A}^{Hall}=-G_n(\bmath{\nabla}\times\bmath{b})\times\bmath{b}
\end{equation}
is a vector quantity, whose curl describes the evolution of the magnetic field under Hall effect and
\begin{equation}
    \bmath{A}^{AD}=[(\bmath{\nabla}\times\bmath{b})\times\bmath{b}-\bmath{\nabla}Y]\times\bmath{b}
    \label{aadeq}
\end{equation}
is a vector quantity, whose curl describes the evolution of the magnetic field under ambipolar diffusion. 

Equation (\ref{bfeq}) takes the simple form 
\begin{equation}
    \frac{\partial\bmath{b}}{\partial\tau} = \bmath{\nabla}\times\bmath{A}\,.
\end{equation}
Using the field decomposition in a poloidal and a toroidal part, equation (\ref{decompotion}), we can easily find that 
\begin{equation}
    \frac{\partial P}{\partial\tau}=r\sin{\theta}A_\phi
\label{Peq}
\end{equation}
and 
\begin{equation}
    \frac{\partial T}{\partial\tau}= \sin{\theta}A_\phi+r\sin{\theta}\frac{\partial A_\theta}{\partial r}-\sin{\theta}\frac{\partial A_r}{\partial\theta}
\label{Teq}
\end{equation}

In our simulations we solve equation (\ref{Yequation}) for the current values of potential $P$ and $T$, then we integrating equations (\ref{Peq}) and (\ref{Teq}) to calculate the evolution of $P$ and $T$ using relations (\ref{aeq}-\ref{aadeq}).

\subsection{External field}

Outside the star we assume a vacuum, thus there are no currents and as a result the external magnetic field is the gradient of a scalar field $\Psi$ such that $\Psi$ is the solution of equation 
\begin{equation}
    \nabla^2\Psi=0
\end{equation}
We decompose the external field in a sum of multipolar expansion \citep{Marchant2011}:
\begin{equation}
    \Psi=\sum^\infty_{l=1}\frac{a_l}{r^{l+1}}P_l(\cos{\theta})
\end{equation}
where the coefficients $a_l$ can be calculated as follows
\begin{equation}
a_l=\frac{2l+1}{2(l+1)}\int^\pi_0 P(R_s,\theta)P^1_l(\cos{\theta})d\theta
\end{equation}
where $R_S$ is the neutron star radius, $P_l(x)$ is the Legendre polynomial of degree $l$ and $P_l^1(x)$ is the associated Legendre polynomial of degree $l$ and order 1. 

\subsection{Boundary conditions}

On the surface of the neutron star the radial field must be continuous, thus the $P$ potential, at $r=R_S$, have to satisfy the condition 
\begin{equation}
\frac{\partial P}{\partial r}=-\sin{\theta}\frac{\partial\Psi}{\partial\theta}
\end{equation}
In addition $B_\phi$ is zero outside of the star, thus we set on the surface of the star $B_\phi=0$ and consequently $T=0$. 

Inside the neutron star, due to axial symmetry no field line is permitted to cross the symmetry axis. This means that $B_\theta=B_\phi=0$ on the axis, or in terms of the  scalar functions $T=0$ and $\partial P/\partial r=0$ respectively. For simplicity we choose that the constant value of $P$ to be zero on the symmetry axis. 

Lastly at the crust - core interface $r=R_c$ we impose the condition that the radial component of ambipolar diffusion velocity to be zero. From equation (\ref{vequation}) this condition means that 
\begin{equation}
    \frac{\partial\Delta\mu}{\partial r}\Big|_{r=R_c}-\frac{f_{mag,r}}{n_c}\Big|_{r=R_c}=0
\end{equation}

\subsection{Timescales} 
Combining the relevant parameters we obtain 
the Ohmic dissipation timescale:  
\begin{equation}
    \tau_{Ohmic}=\frac{4\pi\sigma L^2}{c^2},
    \label{ot}
\end{equation}
with $L$ being the scale of the region the mechanism acts. 

In our simulations we will use Ohmic dissipation both in the crust and the core. The timescales in the two regions will not be of the same order because of difference on the electrical conductivity, as it will be explained further in the next section, and different scales with the crust having a width of  $0.1 R_s$. This leads to a slower magnetic field evolution due to Ohmic dissipation in the core of the neutron star compared to the crust where the conductivity is smaller and thus Ohmic dissipation has a stronger impact on the evolution of the magnetic field in the neutron star evolution. 

The Hall effect has a timescale: 
\begin{equation} 
\tau_{Hall}=\frac{4\pi n_c e L^2}{cB_0}.
\end{equation}
The timescale of the Hall effect has a dependence on the magnetic field strength making it a faster mechanism than Ohmic dissipation and thus responsible for the evolution in the first kyr of the magnetic field evolution in the crust of the neutron stars, provided that the magnetic field is sufficiently strong. For typical values of crustal microphysics this occurs at approximately $10^{13}$G.

With respect to ambipolar diffusion, in a similar manner we can derive from the evolution equation (\ref{Bequation}) the timescale 
\begin{equation}
    \tau_{AD}=\frac{4\pi n_c m_p^* L^2}{x_n^2 t_{pn} B^2}\,. 
\end{equation}
The dependence on the magnetic field is more drastic now making the ambipolar diffusion the dominant mechanism in the core of a highly magnetized neutron star. In ambipolar diffusion timescale there is an implicit dependence on the temperature through the relaxation time of proton-neutron interactions \citep{Yakovlev1990}
\begin{equation}
    \frac{1}{t_{pn}}=4.7\times 10^{16} T_8^2 \Big(\frac{\rho_0}{\rho}\Big)^{1/3}s^{-1}.
\end{equation}

The timescale of the ambipolar diffusion can be distinguished into two timescales for the solenoidal and irrotational part of ambipolar diffusion. The gradient of the departures from chemical equilibrium  as explained in subsection \ref{Sec:Amb_Vel} will balance the irrotational part of the magnetic force and thus reducing the component of the velocity. A timescale can be associated with every part of the velocity. In the work of \cite{Passamonti} they examine how the temperature affects the timescales and they find out that at low temperatures $T<10^9$K and mUrca processes the timescale of the irrotational part is some orders of magnitude larger than that of the solenoidal part. On the other hand at higher temperatures the two parts have almost the same timescale. In our simulation below we are interested in these temperatures ($\sim 10^9$K) and thus we do not examine further such differences.

\section{Simulations}

We solve equations (\ref{Peq}) and (\ref{Teq}) for the poloidal and the toroidal functions numerically on a two dimensional grid using a finite difference method. The grid is constructed so that the neutron star has a crust width of 0.1 of the neutron star radius. It consists of 104 points in radial direction and 102 in the angular direction. The time integration is done using a 3-step Adams-Bashforth method.

In the crust of the neutron star we account for both the Hall and the Ohmic dissipation to take action but not ambipolar diffusion. In the core of the neutron star ambipolar diffusion is the dominant mechanism. 

Equation (\ref{Yequation}) is solved at every step for the values of the potentials at the time of the integration while the other variables are fixed during the field evolution. The electric conductivity has different constant values in the core and the crust of the neutron star. 

We study the evolution for four different initial magnetic field configurations. Namely, two configurations of either a pure poloidal or a toroidal initial field. A configuration of a mixed poloidal-toroidal field, where the toroidal field is located in an extended part of the star, containing equal amounts of energy. Finally, we study a system where the toroidal field is initially confined in the region of closed poloidal field lines, the so-called "twisted torus" configuration. 

In every case the initial conditions for the magnetic filed were constructed using combinations of Ohmic modes, solutions of the field evolution equation where only Ohmic dissipation contribute to the evolution. Thus for the poloidal scalar function we use the functions 
\begin{equation}
P_{nl}=A_{nl}rj_l(k_{nl}r)P_l^1(\cos{\theta})\sin{\theta}
\end{equation}
 and for the toroidal scalar function the functions 
 \begin{equation}
  T_{nl}=A'_{nl}rj_l(k'_{nl}r)P_l^1(\cos{\theta})\sin{\theta}
 \end{equation}
where $j_l$ is the spherical Bessel function of order $l$, $P_l^1$ are the associated Legendre polynomials of degree $l$ and order 1. The values of the coefficients, $k_{nl}$ and $ k^{\prime}_{nl}$, and the amplitudes, $ A_{nl}$ and $A^{\prime}_{nl}$, are set by the boundary conditions and by requiring that $B^2$ has a volume average equal to 1 over the interior of the star respectively. The values used are the ones presented in Tables 1 and 2 of \cite{Castillo2017}.

The neutron star has a radius of $10$km and and the strength of the magnetic field is normalised to $B_0=10^{15}$G. The conductivity and number density of charged particles are chosen to be uniform, in particular $n_c=10^{36}$cm$^{-3}$, $x_n=0.9$, $\sigma_{core}= 8 \times 10^{24} $s$^{-1}$ and $\sigma_{crust}= 1.14 \times 10^{24}$s$^{-1}$,  resulting in a Reynolds number, $R_n=\sigma B_0/ecn_c$, of approximately 80 in the crust. We have to note here that the purpose of the values used in this study is not to try to fit our model in specific set of physical observations but to investigate how the the evolution in crust under the Hall effect and in the core under the ambipolar diffusion are coupled to each other, while ensuring numerical convergence. In this context the values used are selected so the interplay beetween the two mechanism are not obscured very fast by the diffusion of the field. The other variables are $t_{pn}=2.8 \times 10^{-19} $s and $ \lambda = 8.94 \times 10^{33}$ erg$^{-1}$cm$^{-3}$s$^{-1}$  corresponding to a temperature of the neutron star of $10^9$K. The resulting evolution time is $\tau_B \simeq 2.2 Myr$, which is used for scaling throughout this work.

\subsection{Crust - core interface}

The evolution in the crust proceeds under the Hall effect and the Ohmic dissipation. In the core, the evolution is due to ambipolar diffusion and a subdominant Ohmic dissipation. Switching abruptly between the two regimes, leads to discontinuities of the magnetic field evolution which, due the non-linear terms involved, results in computational artifacts. A way to deal with them is to increase the dissipation of the magnetic field by adopting lower values for the electrical conductivity. The physical parameters involved in the magnetic field evolution, namely the electrical conductivity and the density of the charged particles is expected to change with radius in a realistic neutron star. This will result to smoother changes. So, we approximate a region close to the crust-core interface where the physical parameters are not homogeneous. This way, the transition between the different mechanisms is smooth.

We have implementing that, by choosing a zone of small width of $~0.05R_0$ around the crust-core interface. In that zone the field evolves under all mechanisms (Hall, Ohmic and ambipolar diffusion)  but the coefficients of the terms in equation (\ref{bfeq}) are linearly interpolated between the values they have in the crust and in the core. The coefficient of ambipolar diffusion moving outwards goes from  unity to zero, the coefficient of Hall effect starts from zero becoming $G_n$ in the crust and the coefficient of Ohmic dissipation from the value of $D_n$ in the core to the value of $D_n$ in the crust. 

\begin{figure} 
    \includegraphics[width=.46\textwidth]{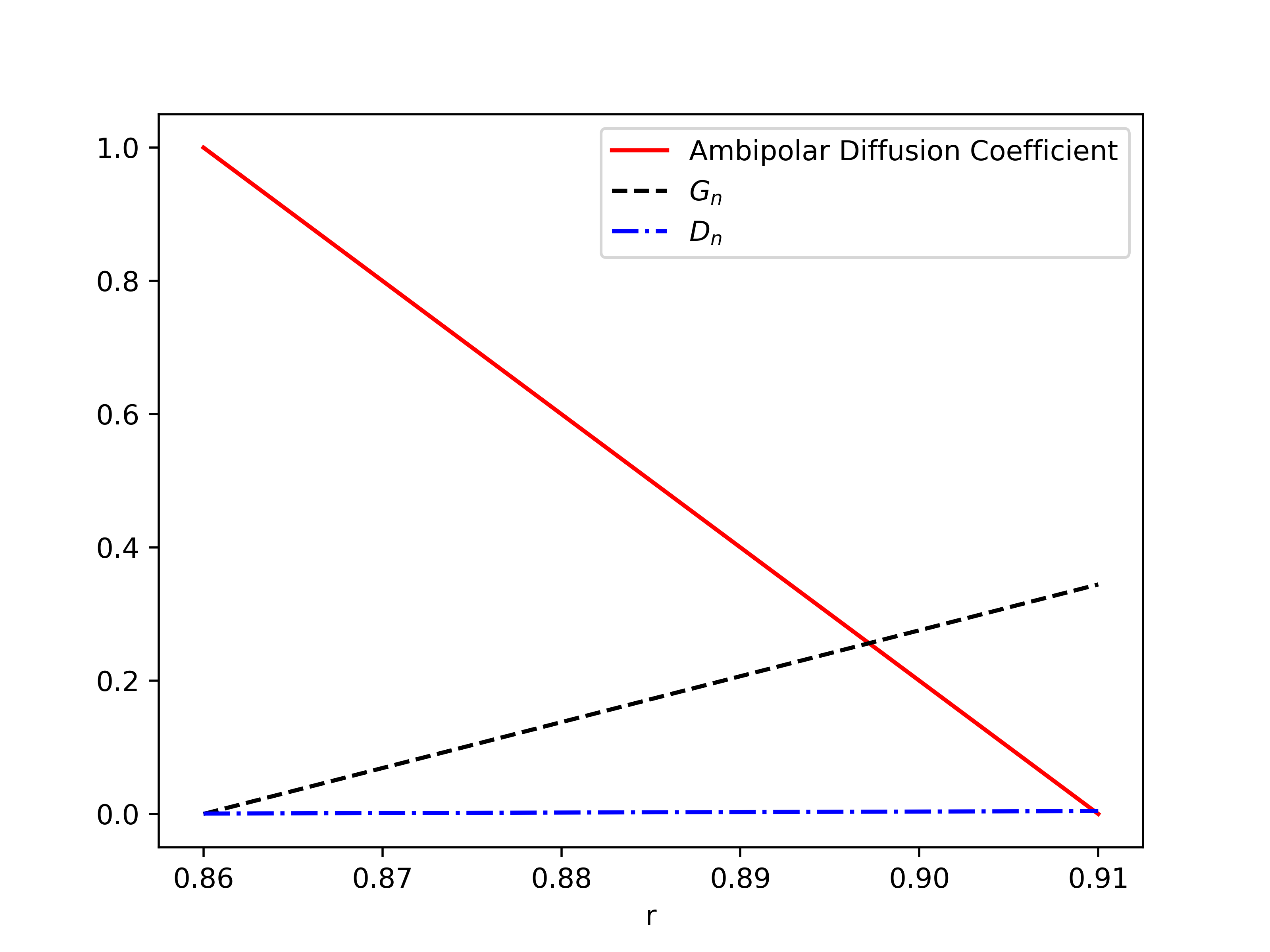}
    \caption{ The values of the coefficients of evolution equation at crust - core interface. We plot with solid line the coefficient of ambipolar diffusion, with dashed line the $G_n$ and with dash-dotted line the $D_n$}
    \label{fig:7}
\end{figure}

The region of the smooth transition between the core and the crust is selected in such a way that the Hall effect is still the dominant mechanism in the crust ($r\geq 0.9 R_0$) (see figure \ref{fig:7}) while ambipolar diffusion dominates the evolution in the core ($r \leq 0.9R_0$).

 \subsection{Purely poloidal initial field}

 We start our study from the case where only a poloidal field exists in the star. The initial poloidal potential is $P(r,\theta)=-P_{11}(r,\theta)/\sqrt{2}$ and the toroidal field is set to be zero, i.e. $T(r,\theta)=0$. 
\begin{figure*} 
    \includegraphics[width=.21\textwidth,height=.30\textwidth]{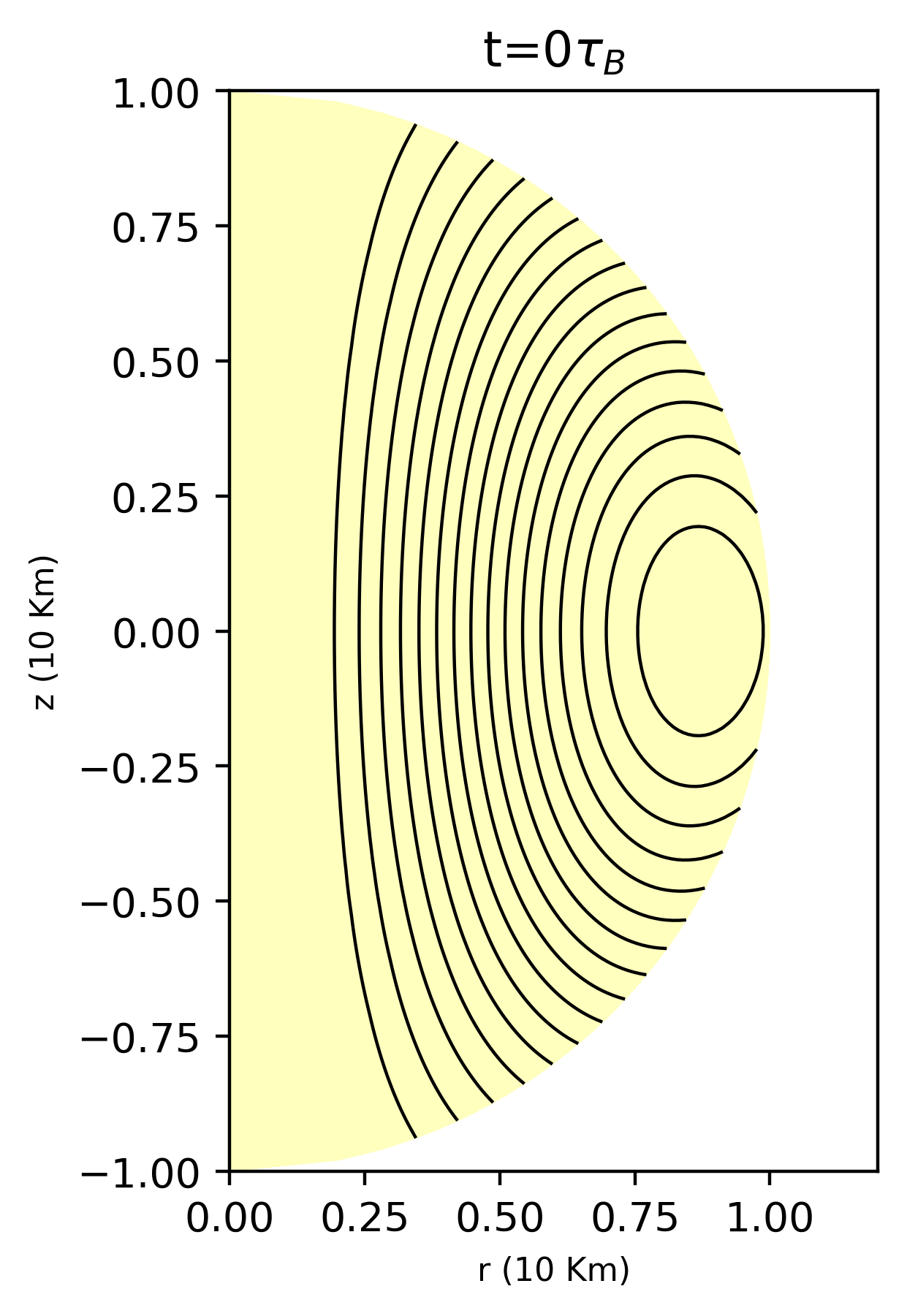}\hfill
    \includegraphics[width=.26\textwidth,height=.30\textwidth]{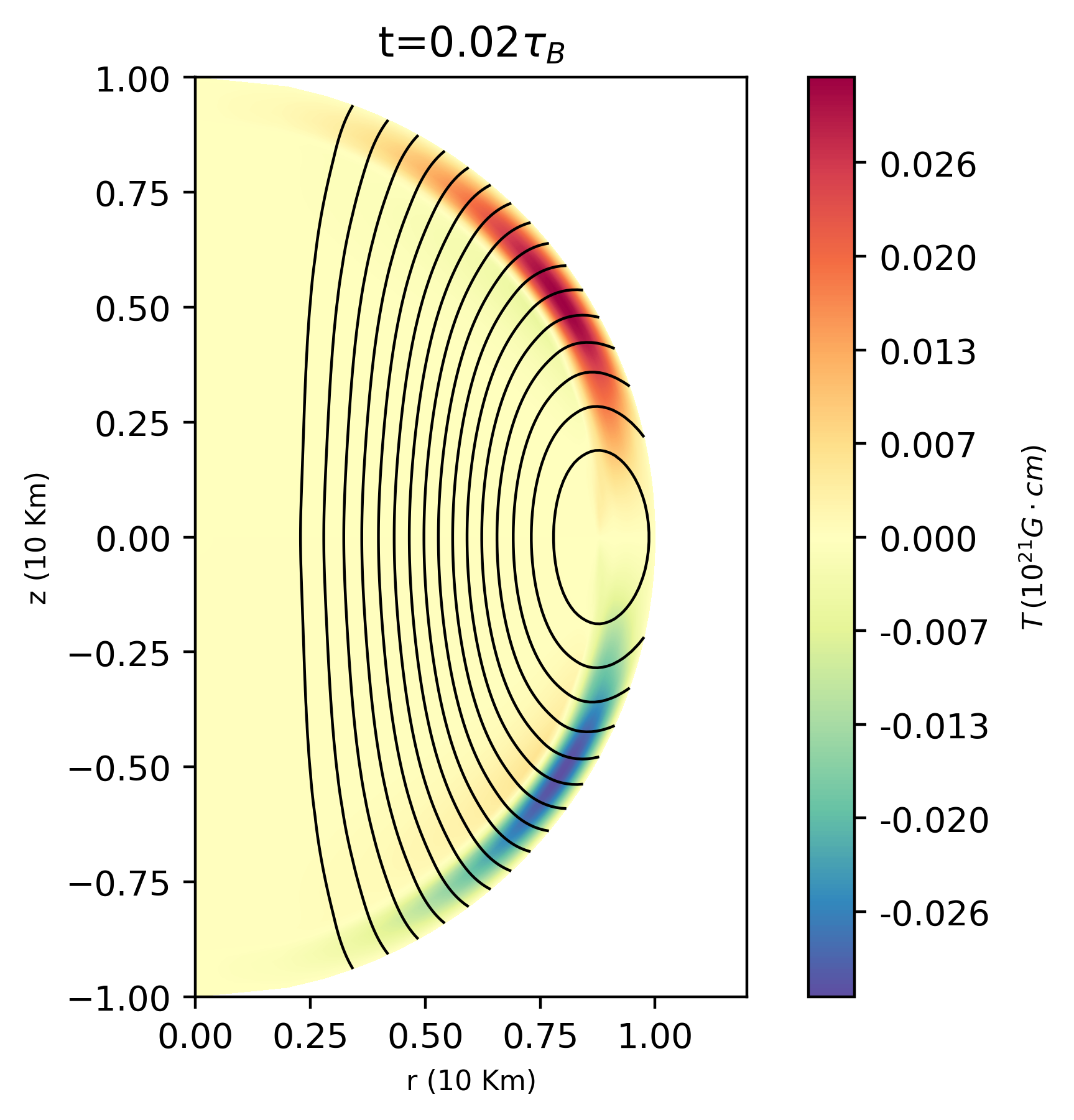}\hfill
    \includegraphics[width=.26\textwidth,height=.30\textwidth]{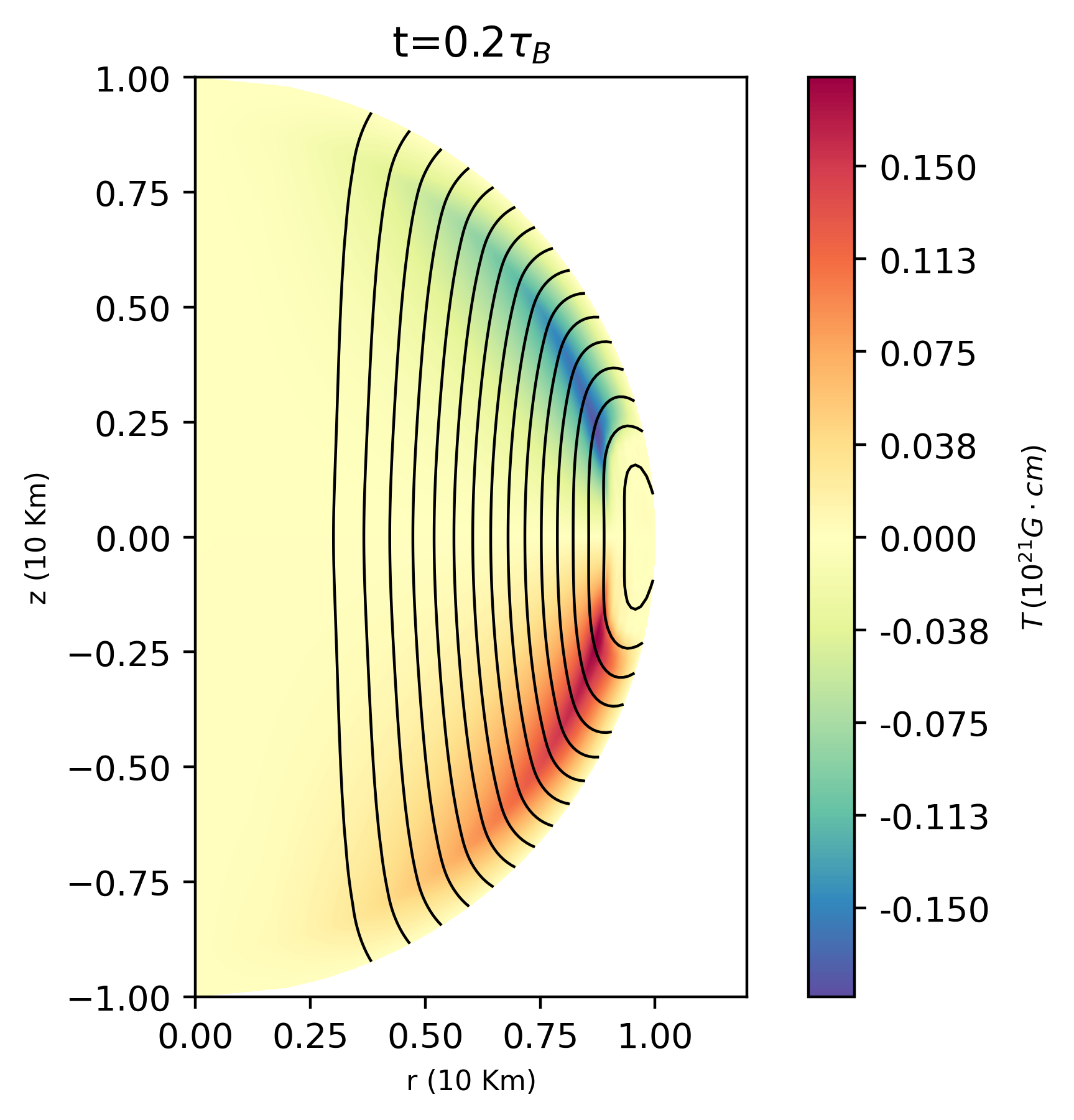}\hfill
    \includegraphics[width=.26\textwidth,height=.30\textwidth]{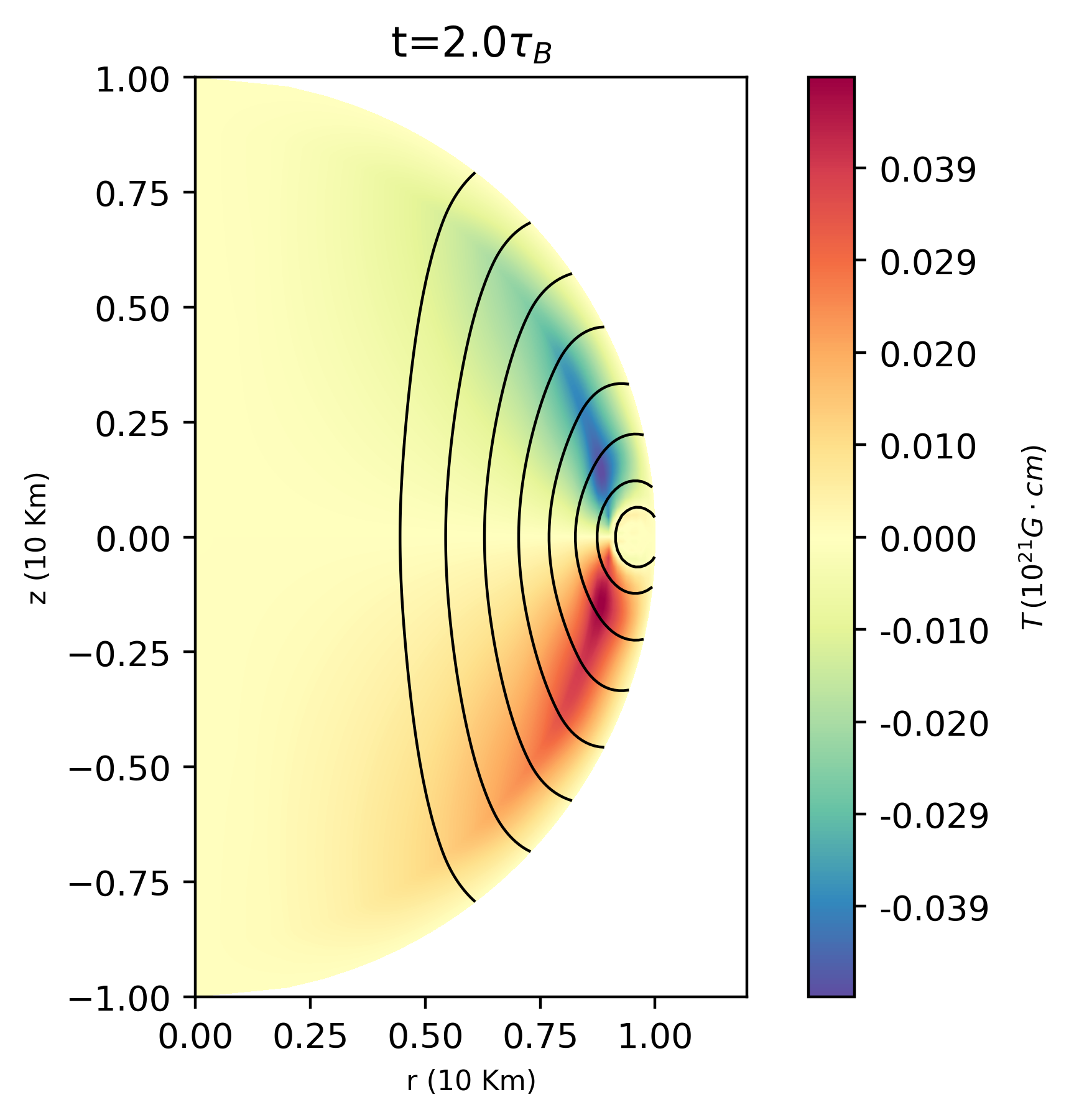}\hfill
    \caption{The magnetic field in four different times, starting from a initially purely poloidal magnetic field. The black lines are the magnetic lines of the poloidal field and color represents the toroidal field function $T$. }
    \label{fig:1}
\end{figure*}

The evolution is presented in figure \ref{fig:1}. At the first few kyrs of evolution the Hall effect generates a toroidal component from the initially purely poloidal field. Ambipolar diffusion pushes the poloidal field lines outwards. As a result the poloidal field lines become  straight. Eventually the field dissipates due to the Ohmic dissipation. The toroidal component relaxes to a quadrupole configuration concentrated primarily at the crust-core interface but also extending in both the core and the crust. The evolution of the crustal field here determines the final state of the evolution.

 In our simulations the poloidal field lines cross the crust-core interface smoothly, without developing kinks or discontinuities. However, there is a difference in the dissipation at the two regions. This was the case even if we do not consider the crust-core smooth transition as described in the previous subsection.

 \subsection{Purely toroidal initial field} 

 Here, the poloidal field field is set to be zero, $P(r,\theta)=0$ while the toroidal field is $T(r,\theta)=-T_{11}(r,\theta)/\sqrt{2}$
\begin{figure*} 
    \includegraphics[width=.25\textwidth]{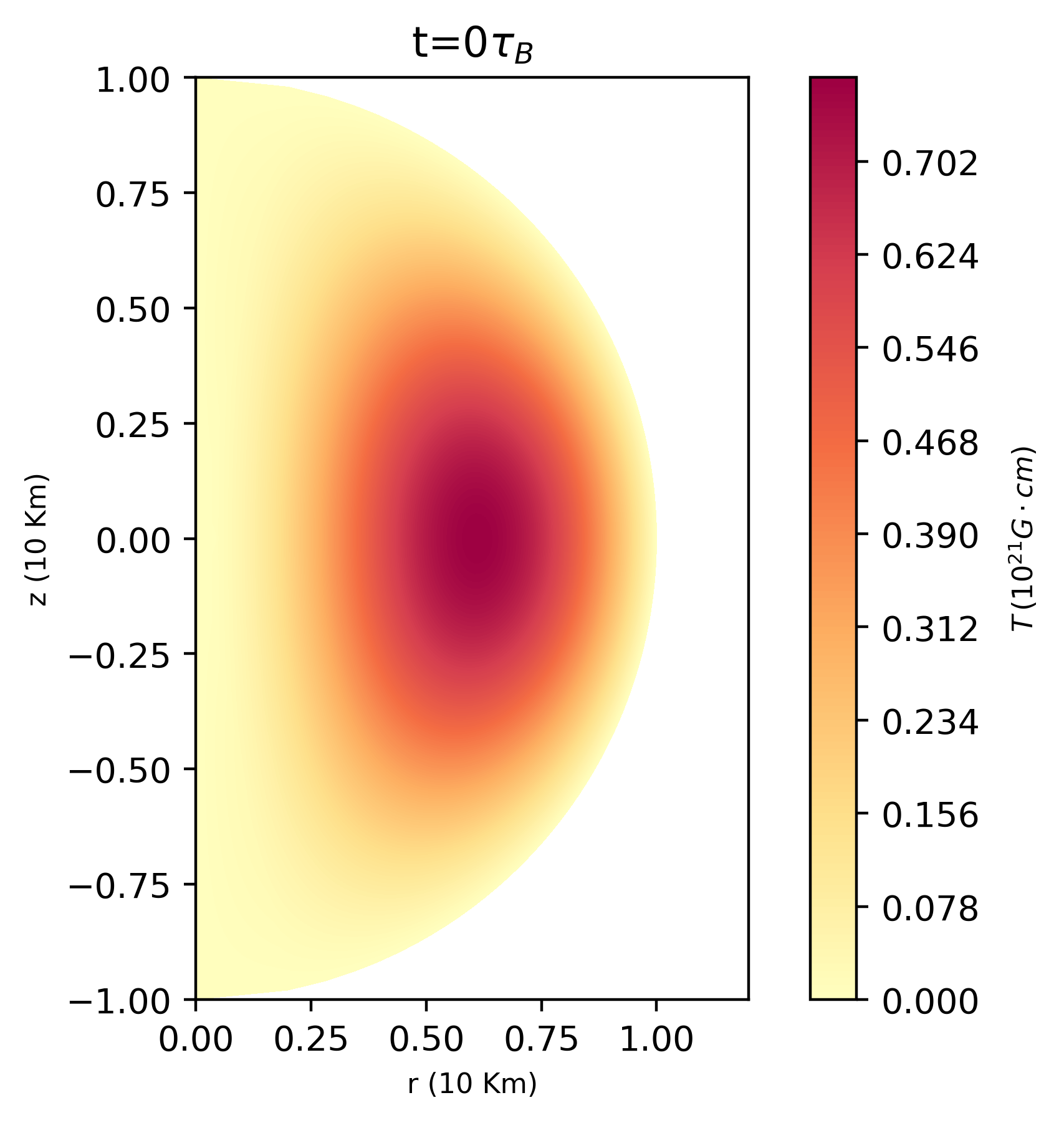}\hfill
    \includegraphics[width=.25\textwidth]{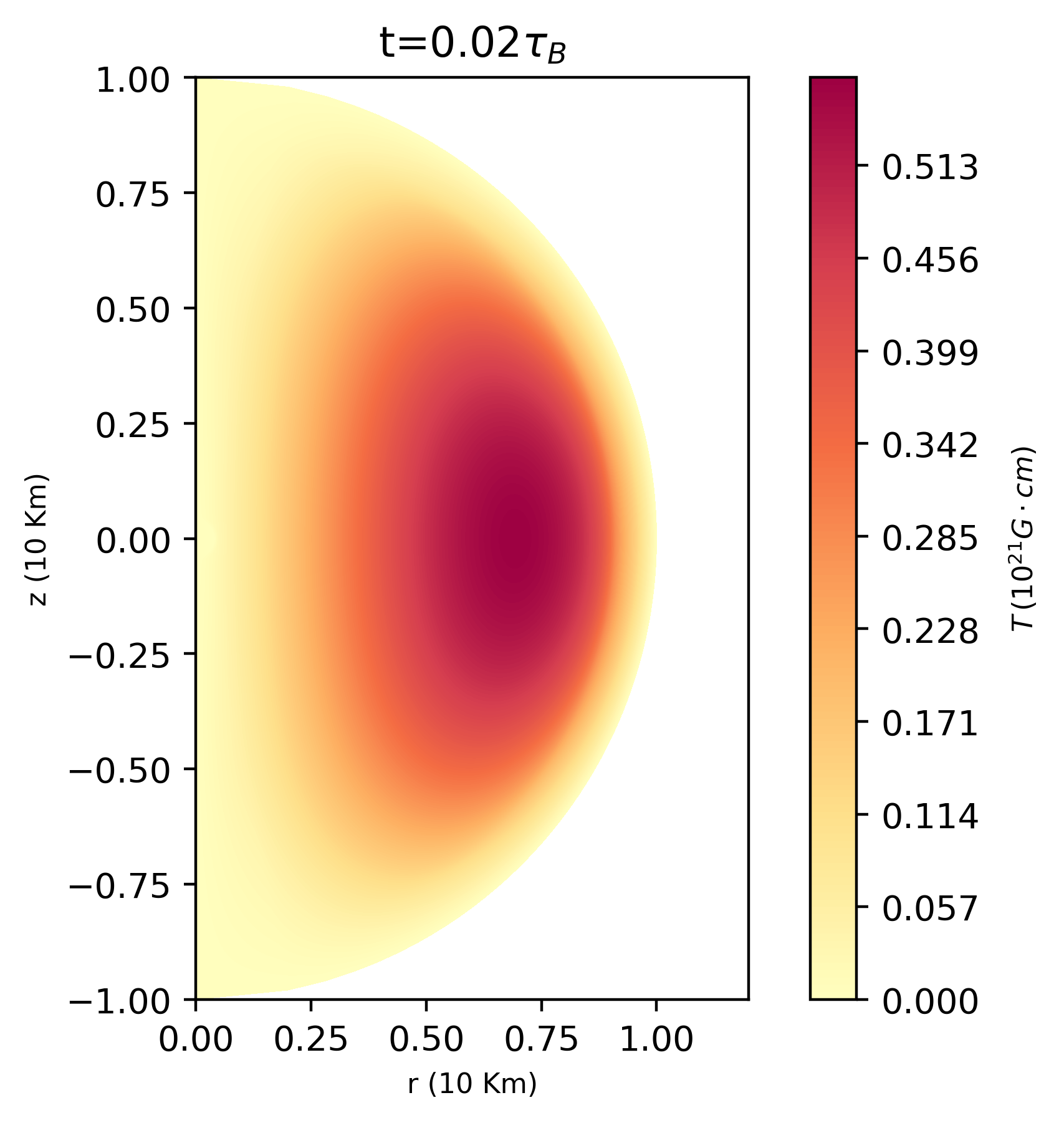}\hfill
    \includegraphics[width=.25\textwidth]{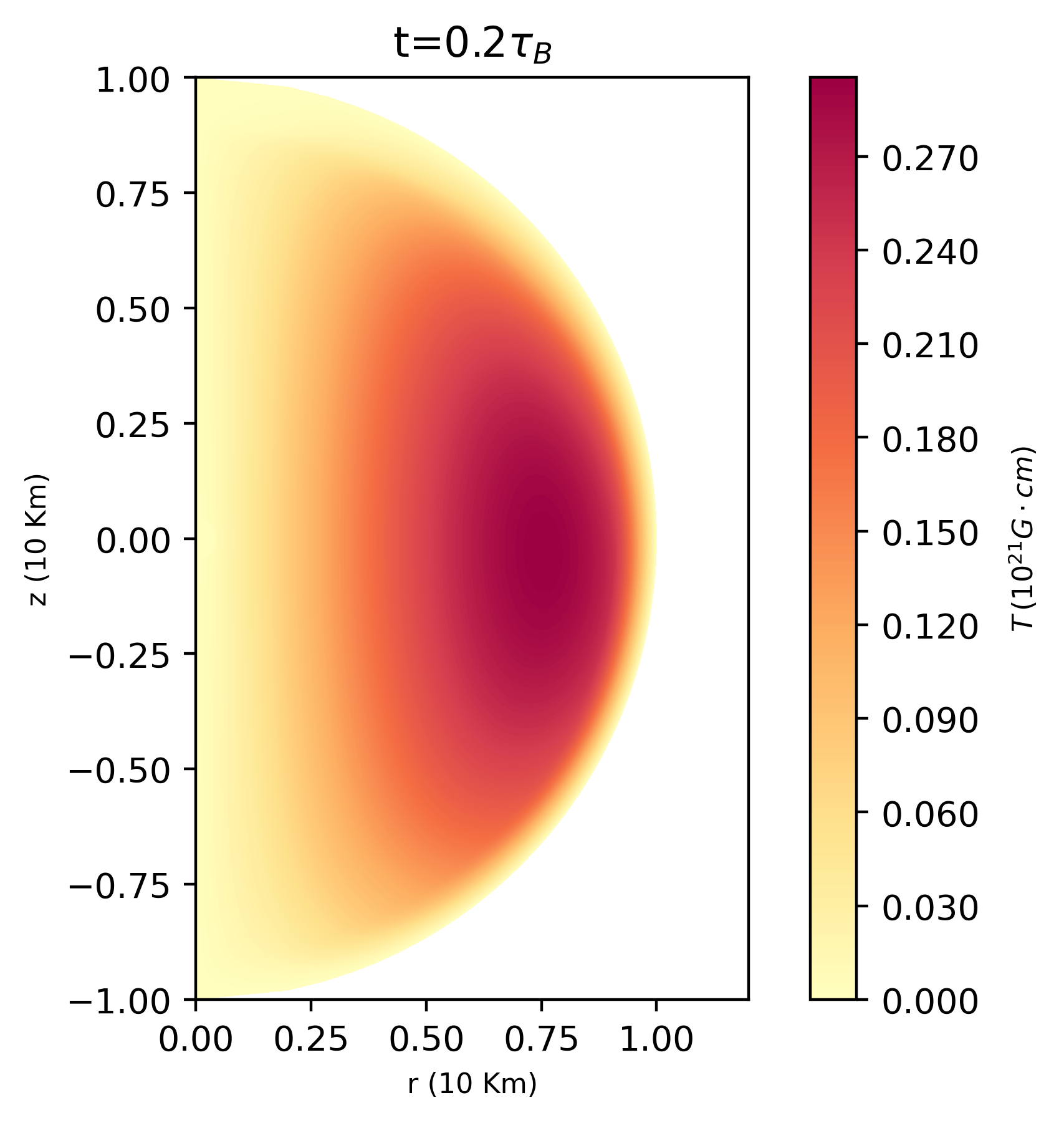}\hfill
    \includegraphics[width=.25\textwidth]{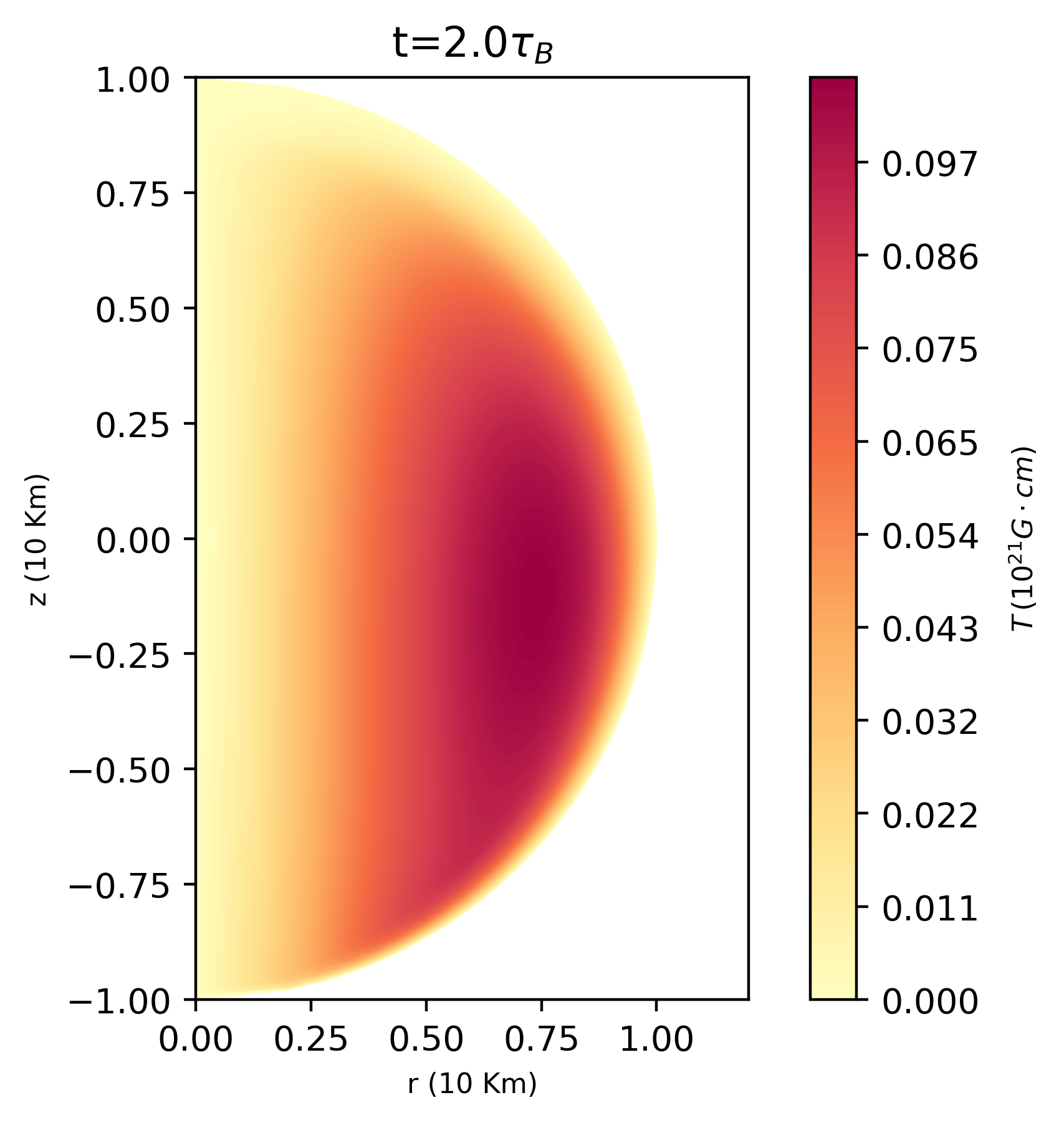}\hfill
    \caption{Snapshots of the evolution of a purely toroidal magnetic field at four different times. The black lines are the magnetic lines of the poloidal field and color represents the toroidal field.}
    \label{fig:2}
\end{figure*}

The evolution of an initially purely toroidal field is simple. Without poloidal field lines to constrain it, ambipolar diffusion pushes the field at the crust-core interface. Meanwhile the field in the core decays. In the simulation, due to this decay, the toroidal field is seen as expanding. 

Due to the coupled evolution as the ambipolar diffusion pushes the field in the core towards the interface, there, due to the Hall effect it starts to move southwards creating the configuration shown in the last panel of figure \ref{fig:2}. 

Contrary to the case of a purely poloidal field, if we do not consider a smooth transition at the crust-core interface, the toroidal field will develop discontinues because of the different mechanisms operating on either side of the boundary. Thus, to prevent such discontinuities of the toroidal field the inclusion of a  transition layer there at the crust-core boundary is necessary.

 \subsection{Mixed initial field}
 For the mixed initial field case we choose $P(r,\theta)=-P_{11}/\sqrt{2}$ and $T(r,\theta)=-T_{11}/\sqrt{2}$.
 \begin{figure*} 
    \includegraphics[width=.25\textwidth]{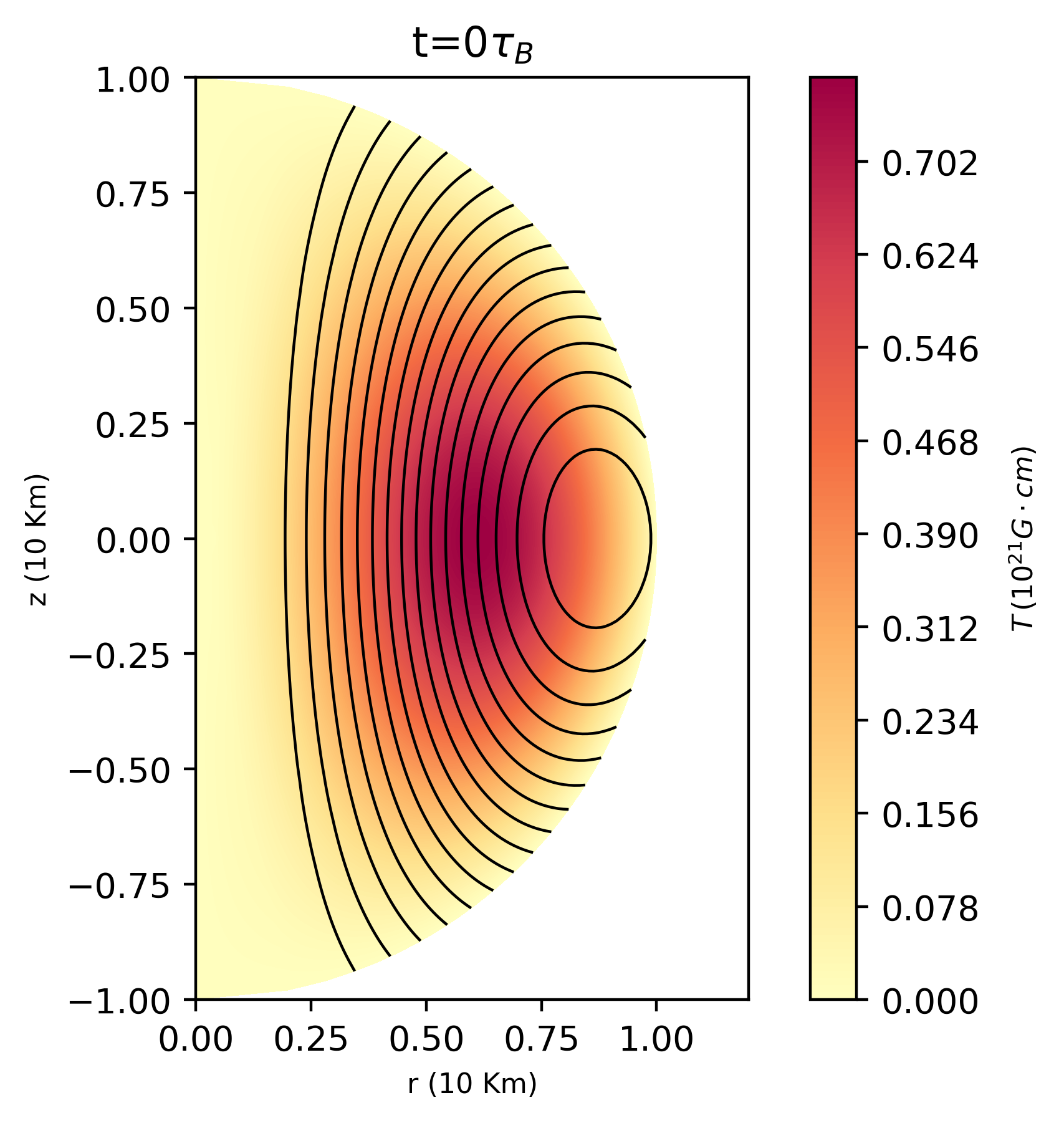}\hfill
    \includegraphics[width=.25\textwidth]{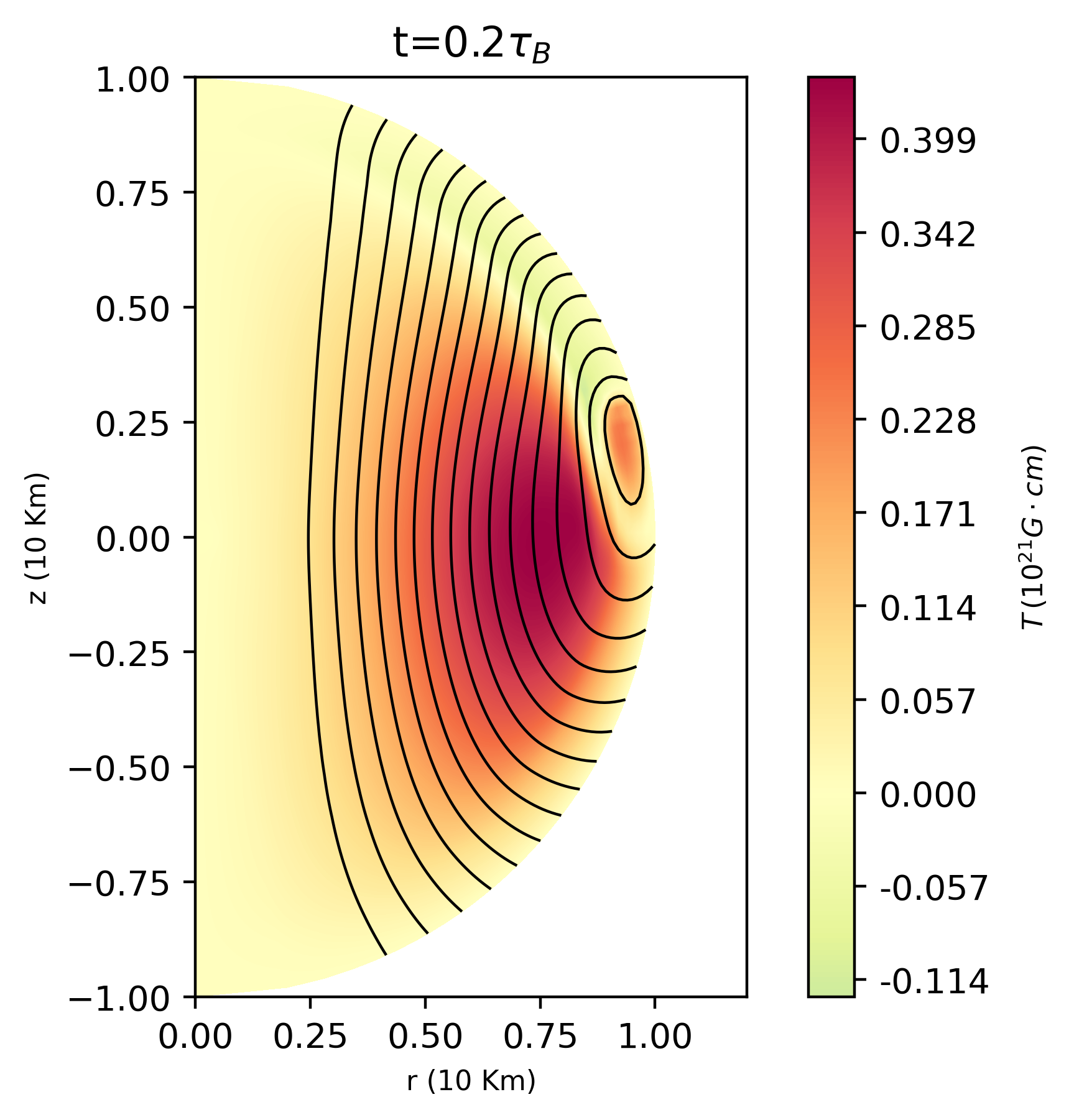}\hfill
    \includegraphics[width=.25\textwidth]{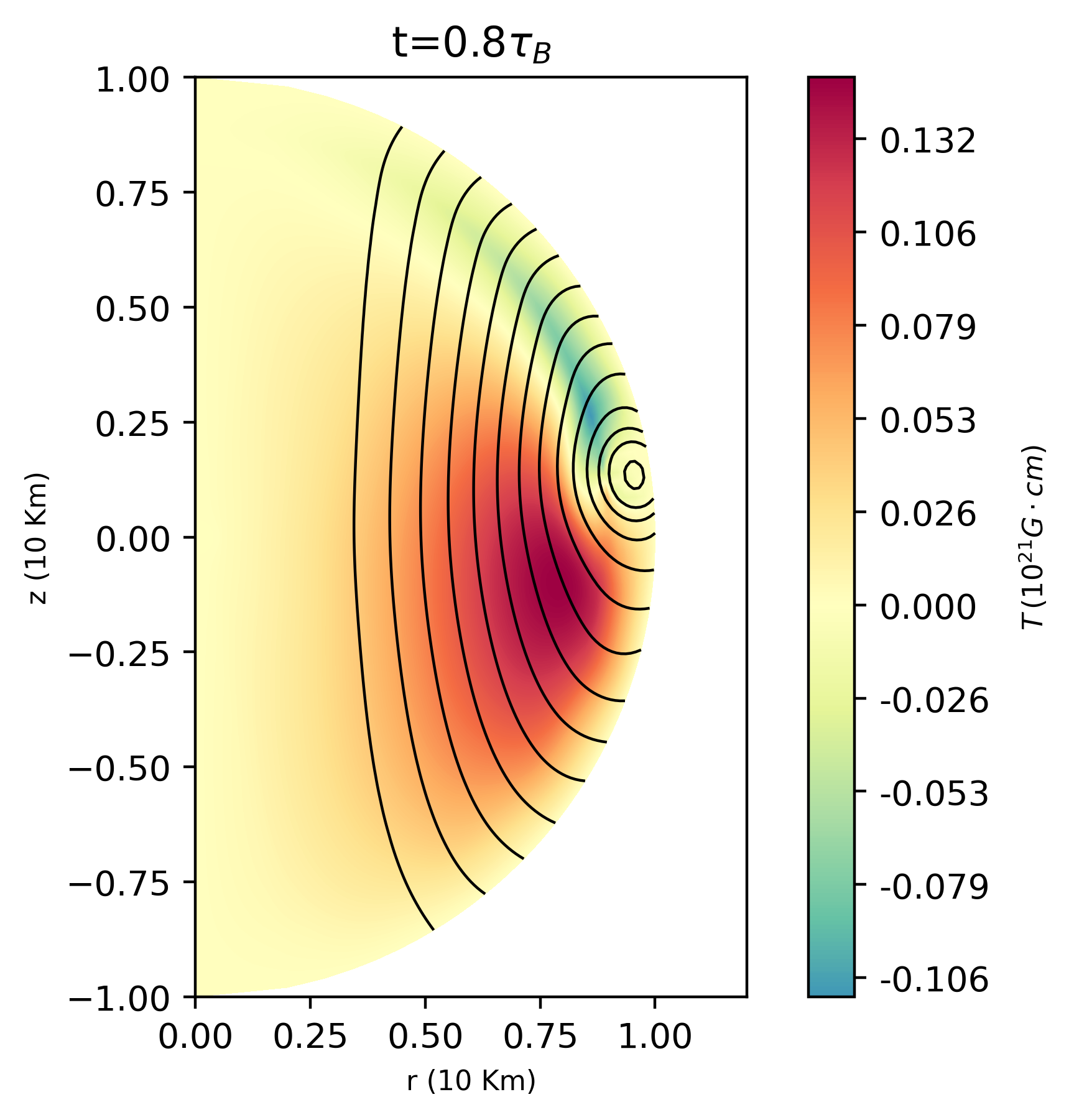}\hfill
    \includegraphics[width=.25\textwidth]{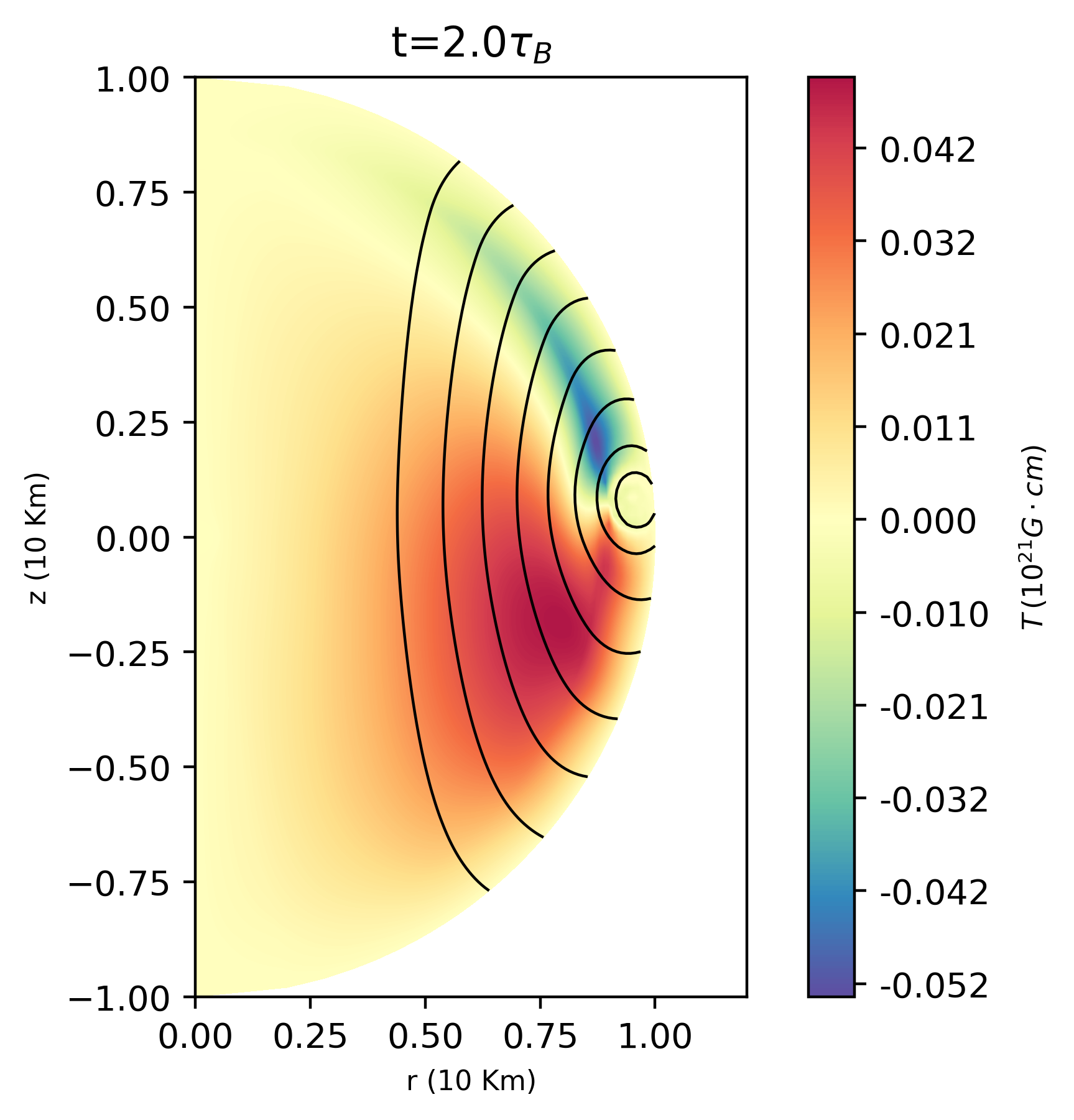}\hfill
    \caption{Instances of the evolution of the magnetic field of the mix initial configuration.  The black lines are the magnetic lines of the poloidal field and color represents the toroidal field function $T$.}
    \label{fig:3}
\end{figure*} 
In figure \ref{fig:3} we show the evolutionary stages of this system. Initially, the toroidal field is pushed towards the crust-core interface and especially where the region of the poloidal field lines that close inside the star. Simultaneously the Hall effect has moved the field towards the northern hemisphere, while some toroidal field is generated in the crust. As a result a part of the initial toroidal field is trapped inside the core and slowly dissipates due to Ohmic decay. 
 
Furthermore, as shown in figure \ref{fig:3}, while the Hall effect breaks the north-south symmetry by pushing the closed poloidal field lines towards the north (south if the initial field has opposite polarity), ambipolar diffusion in the core restores the north-south symmetry at later stages of the evolution. In fact, choosing different values of the physical parameteres  changes the relative strength of the mechanisms, thus a different degree of violation of north-south symmetry in the poloidal field lines is observed.

Eventually the field will relax to a state similar to the pure poloidal field case but with a significant portion of the initial toroidal field existing in the core dissipating slowly due to the mild Ohmic dissipation. Moreover similar to case of the purely initial toroidal field the remaining field in the core is slowly moving southward.

 \subsection{Toroidal field confined in closed poloidal field lines} 

 In this case the toroidal field is everywhere zero except the region of close poloidal field lines. In order to achieve this the toroidal field potential $T(r,\theta)=0$ if $P(r,\theta)<P_0$ where $P_0$ is the surface poloidal potential on the equator, and $T(r,\theta)=s_1(P(r,\theta)-P_0)^{s_2}$ if $P(r,\theta)\geq P_0$ where $s_1$ is a coefficient controlling the relevant strength of toroidal and poloidal field while $s_2$ controls the region where toroidal field is different from zero. In the following we have used $s_1=60$ and $s_2=1.05$. The poloidal field function is $P(r,\theta)=-P_{11}(r,\theta)/\sqrt{2}$

 \begin{figure*} 
    \includegraphics[width=.25\textwidth]{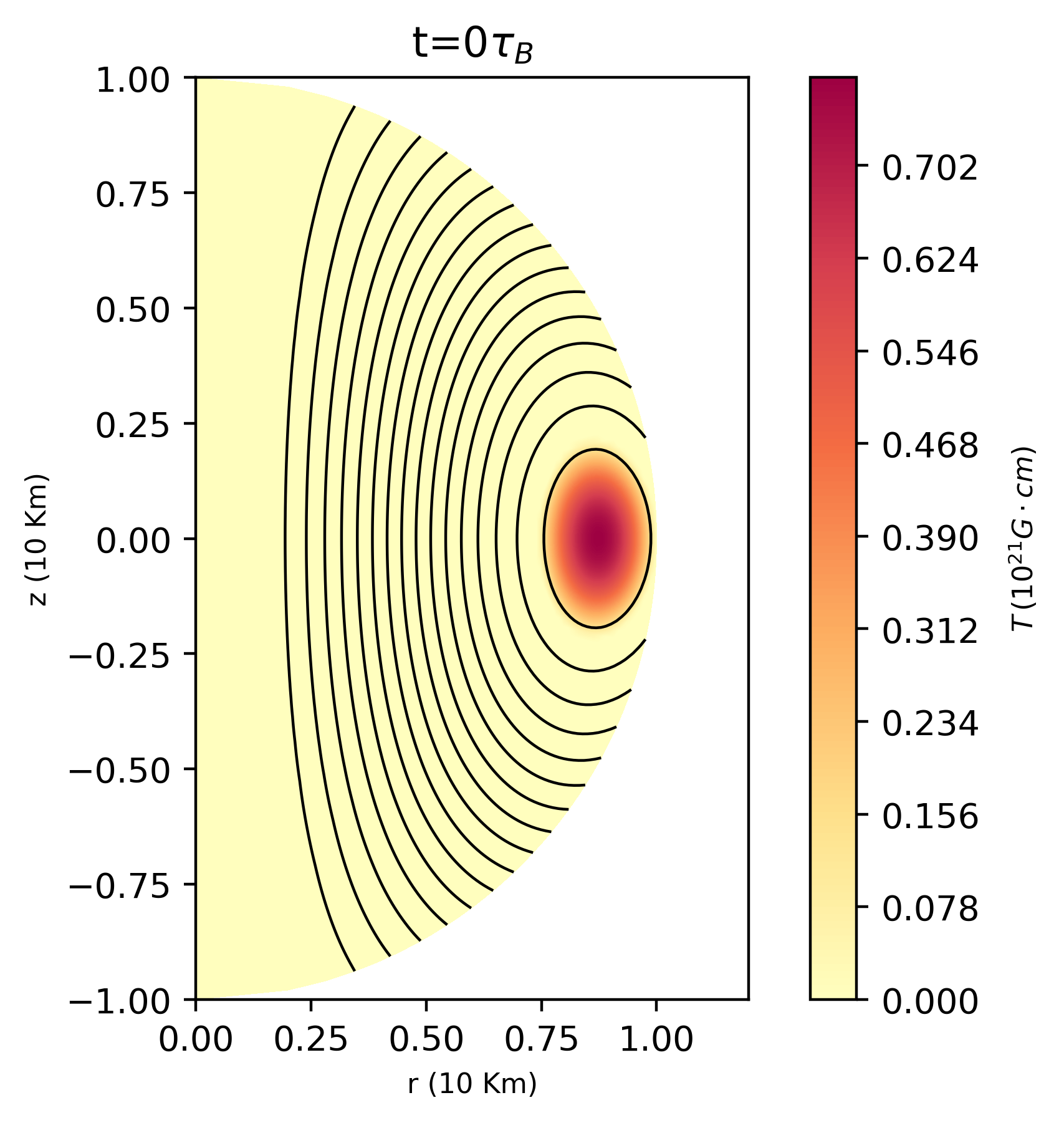}\hfill
    \includegraphics[width=.25\textwidth]{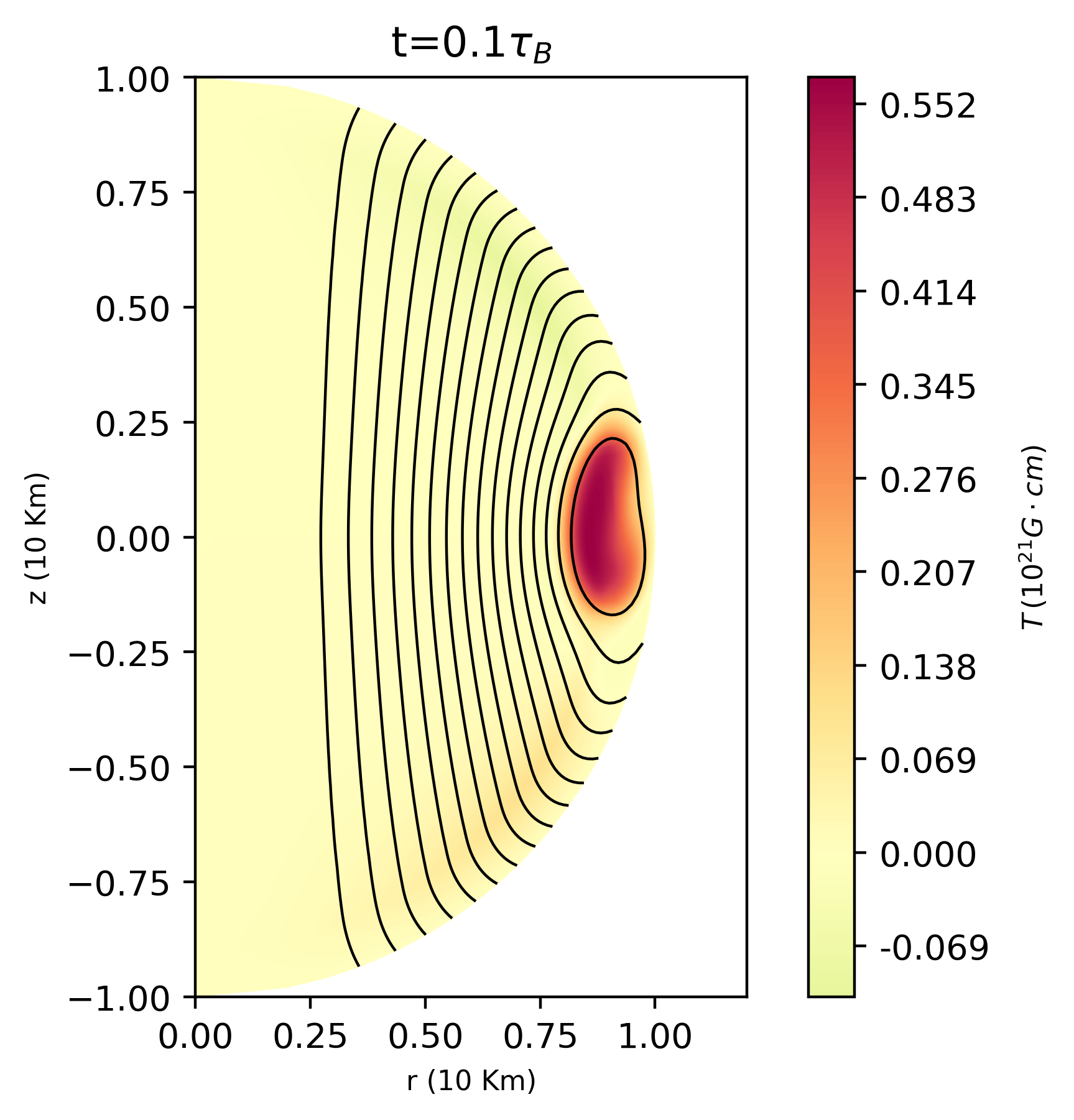}\hfill
    \includegraphics[width=.25\textwidth]{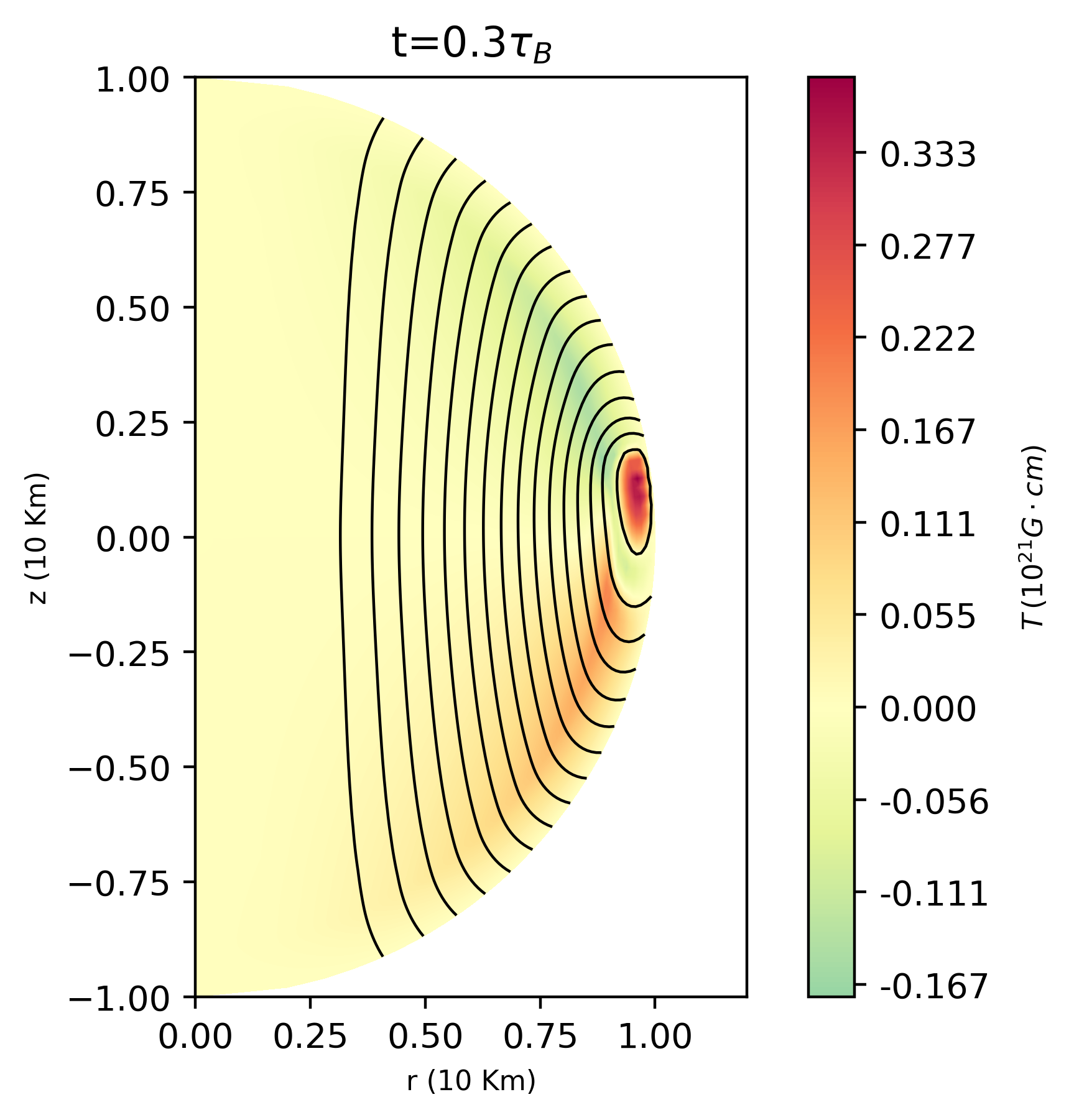}\hfill
    \includegraphics[width=.25\textwidth]{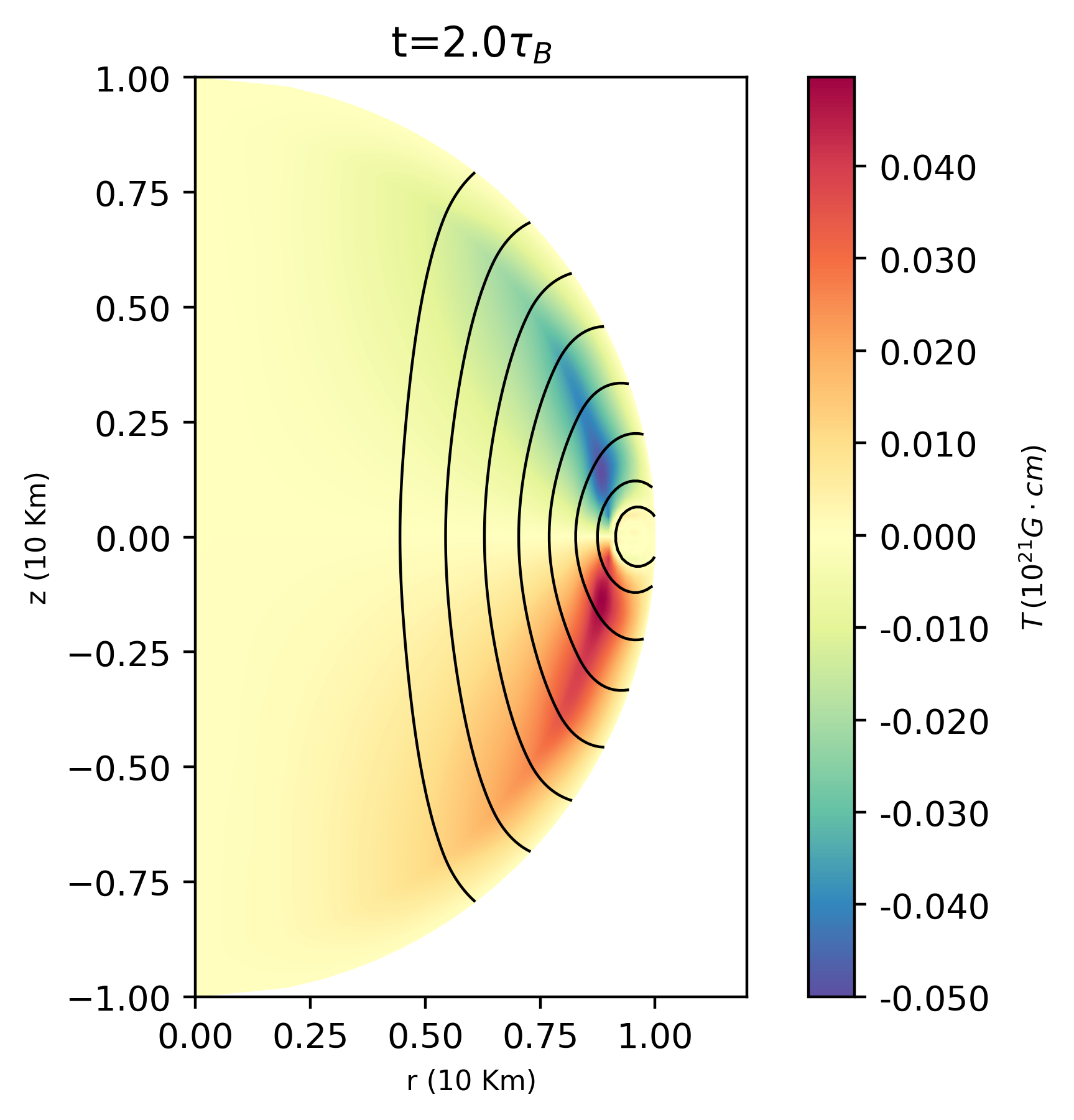}\hfill
    \caption{Snapshots of the evolution of the magnetic field starting form a "twisted torus" configuration. The black lines are the magnetic lines of the poloidal field and color represents the toroidal field function $T$.}
    \label{fig:4}
\end{figure*}

The evolution in this case follows the behaviour of the mixed field configuration. Here the energy of the toroidal and the poloidal field are not equal, as the poloidal field has significantly more energy.  

Thus, the toroidal field in the crust moves north as was the case for the mixed field configuration. However, now the toroidal field that was generated from the Hall effect quickly attains the same strength as the initial toroidal field and eventually overtakes it. Thus, from that point onwards, shown in the  third snapshot of figure \ref{fig:4}, the evolution is similar to the purely poloidal field. The broken north-south symmetry is restored as the dissipation of the magnetic field proceeds in later times. 

\subsection{Magnetic field Energy} 

\begin{figure*} 
    \includegraphics[width=.50\textwidth]{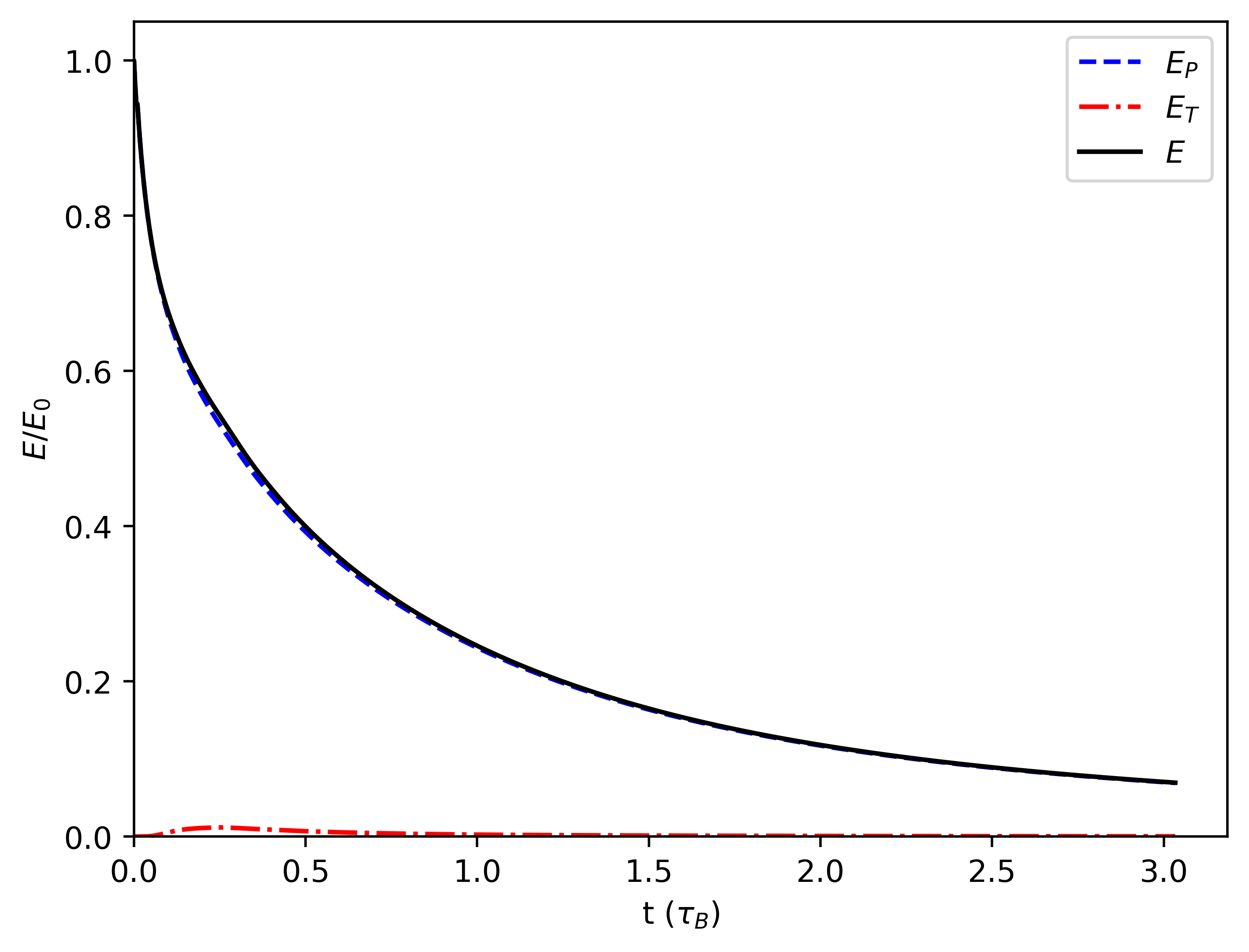}\hfill
    \includegraphics[width=.50\textwidth]{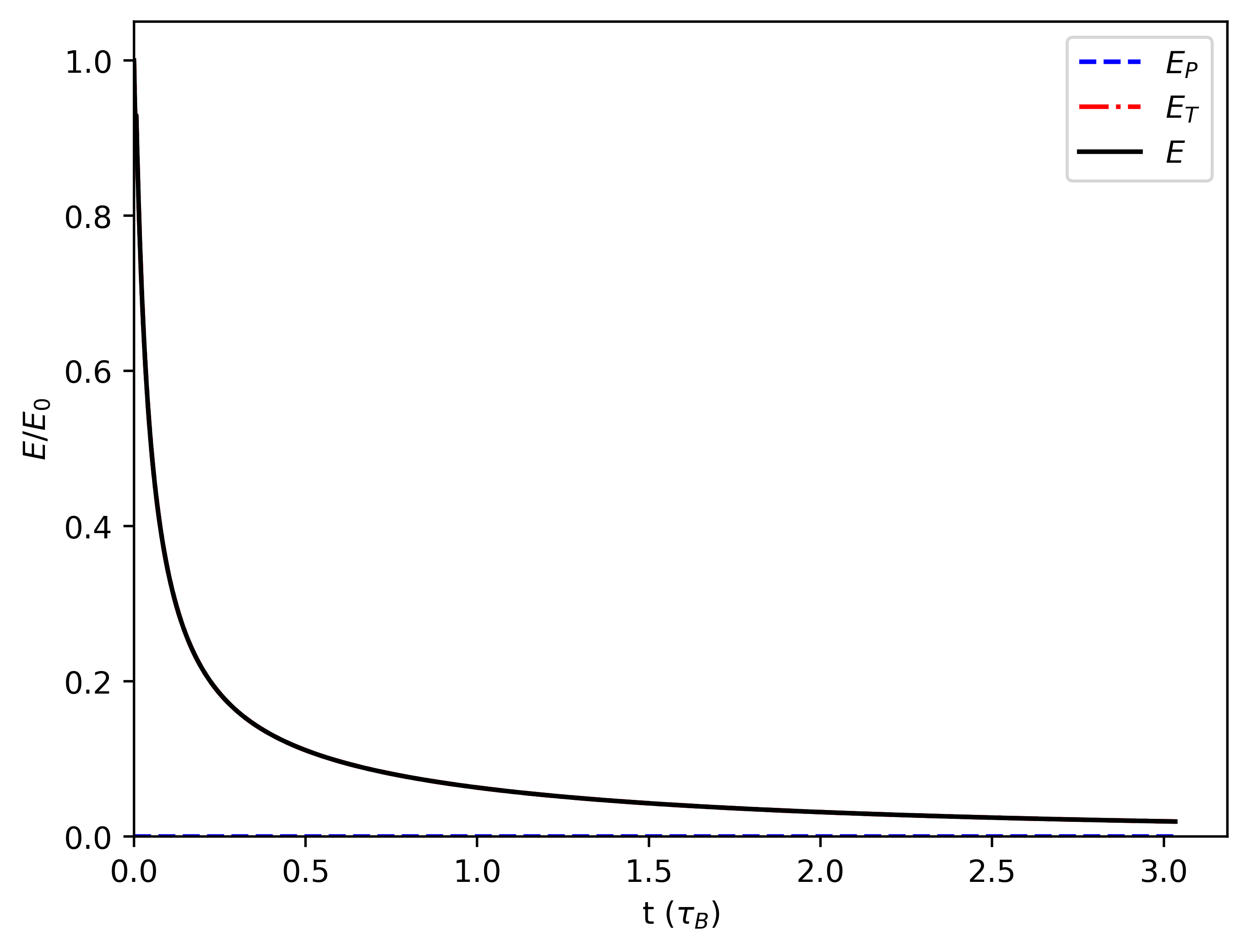}\hfill
    \includegraphics[width=.50\textwidth]{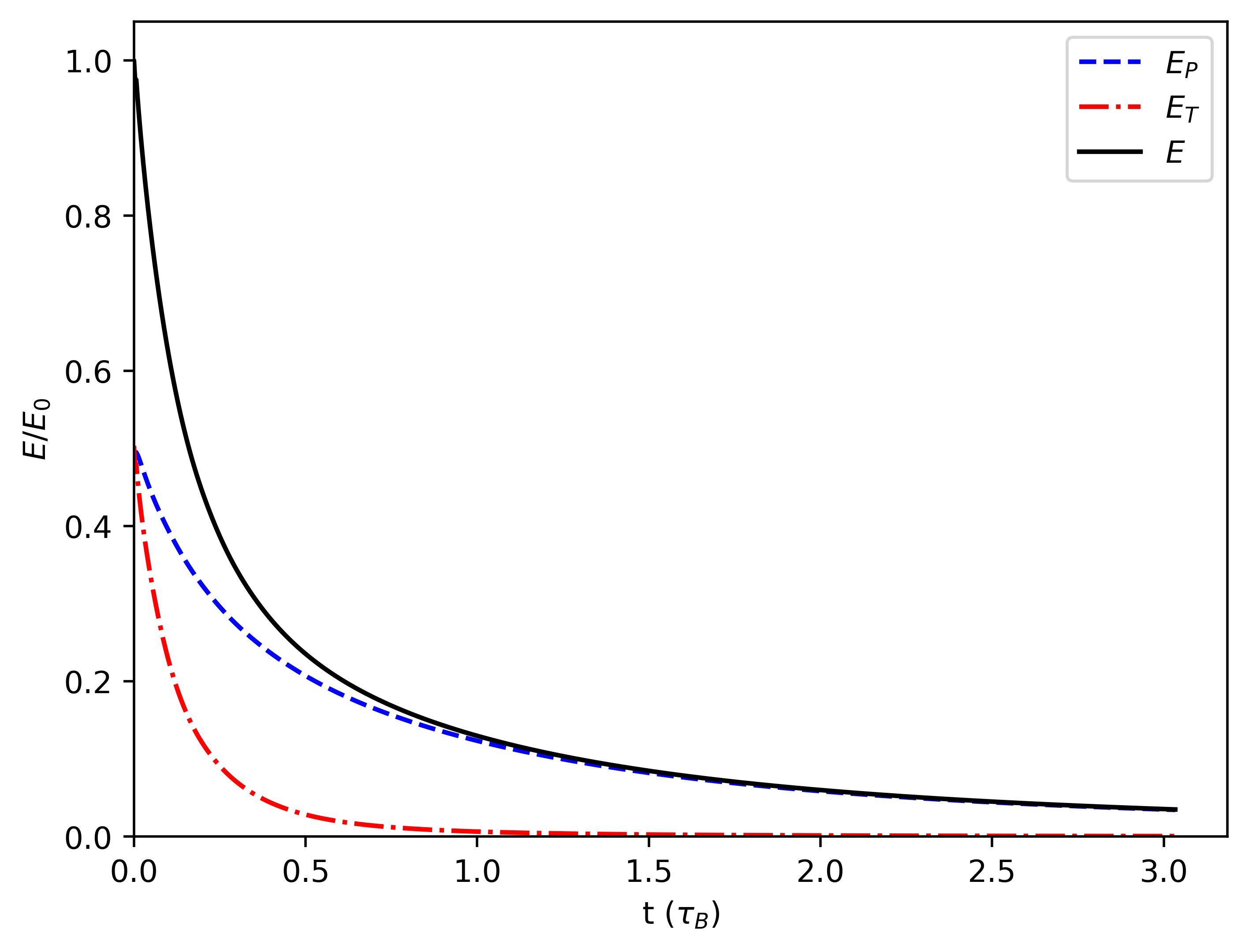}\hfill
    \includegraphics[width=.50\textwidth]{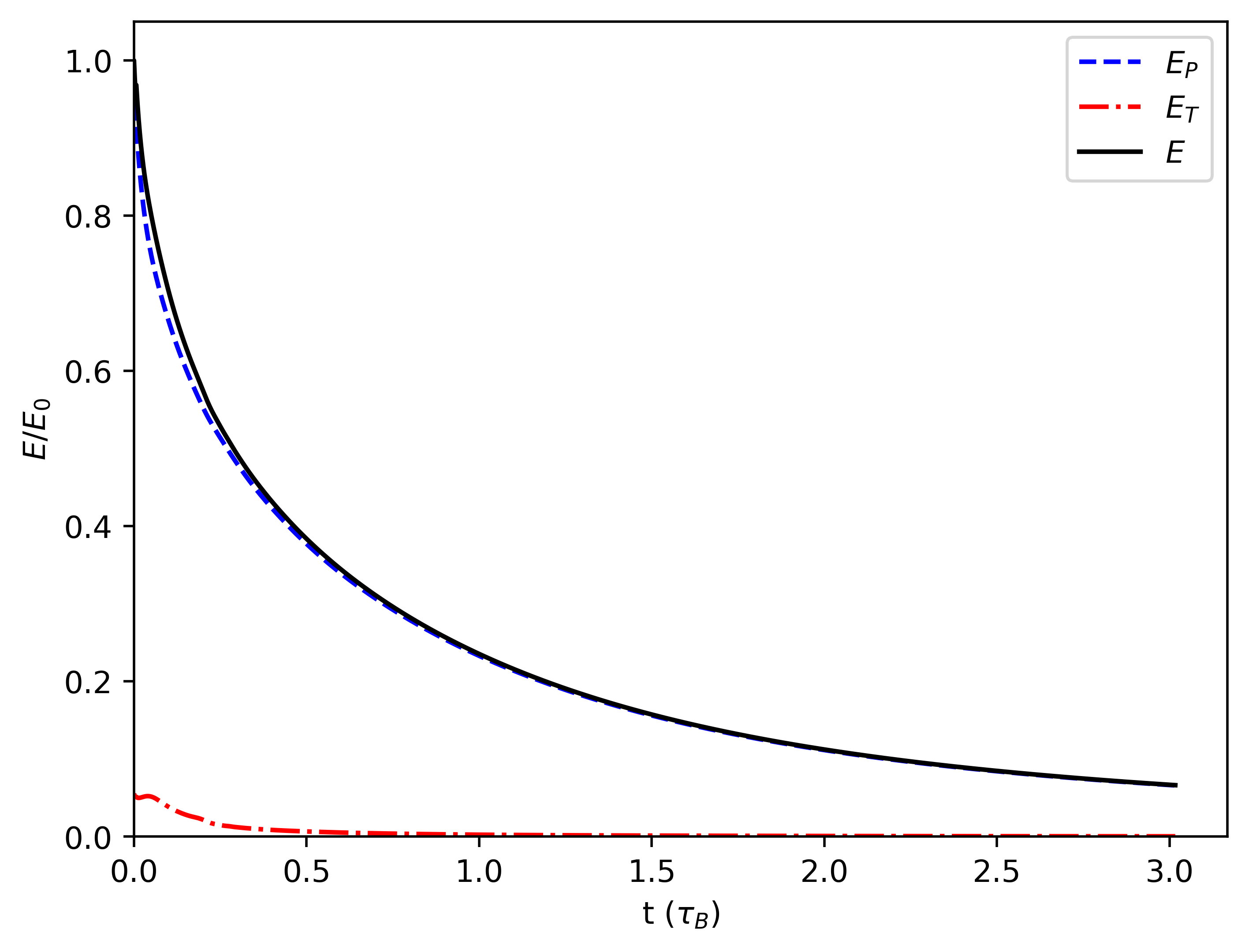}\hfill
    \caption{The energy evolution of the cases studied. On the upper row, the energy evolution of the purely poloidal initial field, left and for the purely toroidal initial field configuration right. At the bottom row the energy of the mixed field with similar energies in toroidal and poloidal component, left and the for the configuration where the low energy toroidal field is initially confined in the region of closed poloidal field lines, right. We plot with solid line the total energy in the magnetic field, with dashed line the energy in the poloidal field and with dash-dotted line the energy in the toroidal field.   }
    \label{fig:5}
\end{figure*}

The energy contained in the magnetic field is calculated  by the integral: 
\begin{equation}
    E_B=\frac{1}{8\pi} \int_V (B_T^2+B_P^2)dV
\end{equation}
where $V$ is the volume of the neutron star. As presented in figure \ref{fig:5} for all the cases considered, there is a fast energy dissipation in the first few kyr of the magnetic field evolution. 

\begin{figure} 
    \includegraphics[width=.46\textwidth]{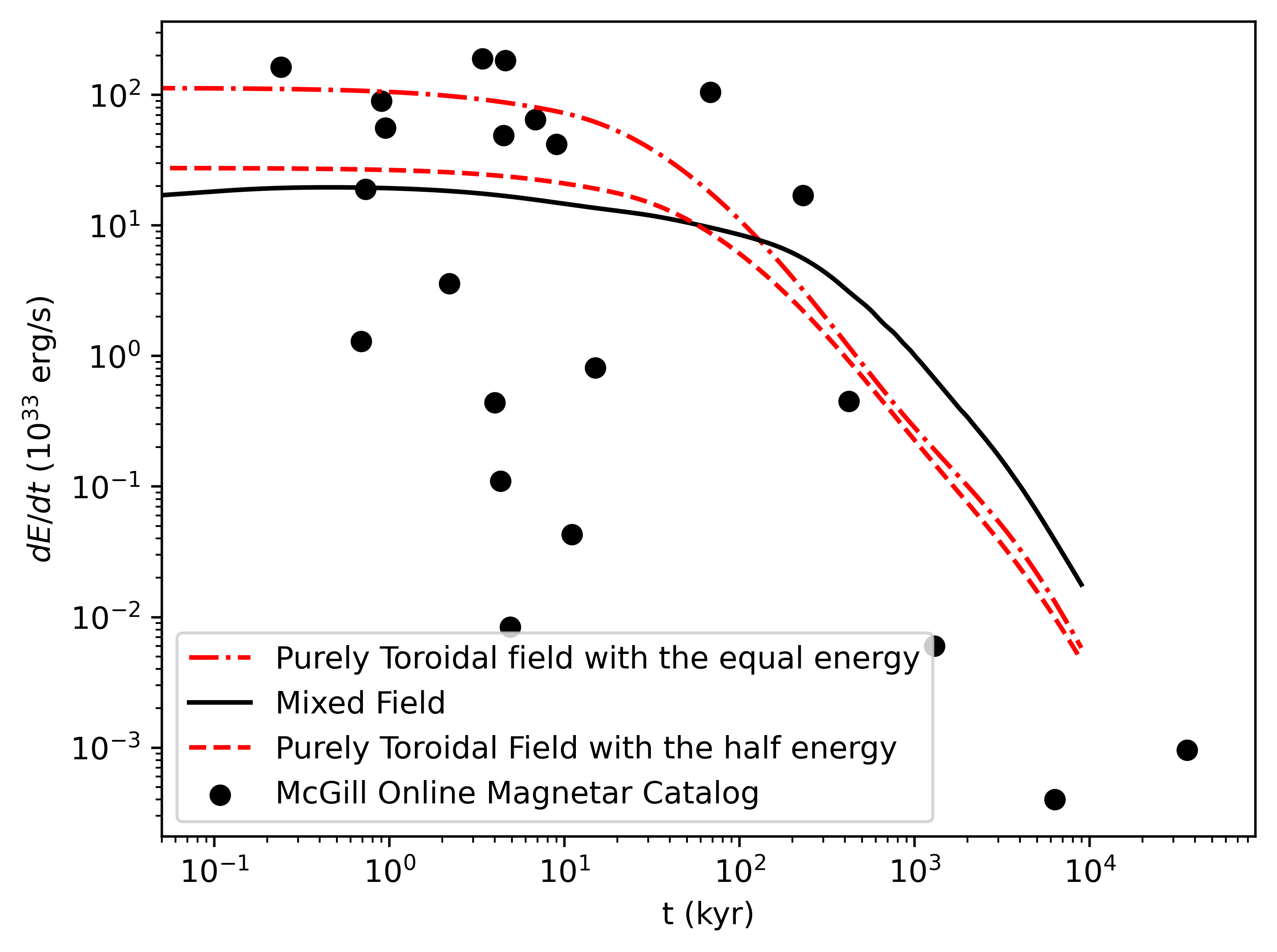}
    \caption{ The energy loss rate of the magnetic field for the case of the initially mixed field configuration (solid) and the cases of a purely toroidal field (dashed lines). We have plotted the energy loss rate of an initial purely toroidal field with initial total energy equal only to the energy of the toroidal component of the mixed field case (dashed line) and with dash-dotted line the energy loss rate of an initial purely toroidal field with initial energy equal to the total energy of the mixed field case. The black points are the X-ray luminosities observed from Magnetars of the  McGill Catalog.}
    \label{fig:6}
\end{figure}

In figure \ref{fig:6} the solid line represents the energy loss rate of the initial mixed field configuration. 
Moreover the energy loss rate in the case of the purely toroidal initial field is also plotted. It is evident that in this case the magnetic field energy dissipates faster. In both cases there is the same initial energy in the toroidal field and the only difference is that in the case of the mixed field a poloidal component with similar energy is also present. The initial poloidal field results in an evolution with initially lower magnetic energy loss. However before the first 100 kyr, in the case of the purely toroidal field the energy loss rate is starting to decrease and becomes slower than that of an initial mixed field configuration in which the drop is observed at later times. 

This behaviour becomes more clear if we consider a case of a purely toroidal initial configuration with energy similar to the total energy of the mixed field configuration (dash-dotted line in figure \ref{fig:6}). There, the energy loss rate is even higher at the first kyr of the evolution and with an earlier time of drop.

In the same diagram, we plot the X-ray luminosity observed from magnetars of McGill Catalog are plotted \citep{2014ApJS..212....6O}.
The cases where initially there is more energy in the toroidal field seem to fit them better further supporting the hypothesis that strong toroidal fields might present in magnetars \cite{Igoshev2021}.

We have to note here that our result are based on the assumption that the baryon velocity is neglected. This assumption results in slower evolution and thus lower energy loss rate.

\section{Discussion}

Our simulations combine the evolution of the core dominated by ambipolar diffusion and the crust dominated by the Hall effect. While the combined evolution is complex, we can identify some trends known from previous works that have focused on either the Hall effect or ambipolar diffusion. 

In the case of the purely toroidal field and constant crust density the magnetic field will be displaced due to the Hall effect in the crust, in $\textbf{z}$ direction. A positive field as the one we simulate here will move in the $-\textbf{z}$ direction, figure \ref{fig:2}. In the case of the purely poloidal field the Hall effect will generate toroidal and rearrange the energy of the field components. 
Furthermore, a crust-confined magnetic field with a mixed field configuration, that evolves due to the Hall effect
will be driven towards a pole depending on the polarity of the initial field, breaking the north south symmetry of the configurations. For a positive polarity initial configuration there will be movement towards the north while for a field with negative polarity towards the south  \citep{Reisenegger2007etal,Pons2009,Vigano2013, Gourgouliatos2013}. This is indeed what we see at the first stages of the magnetic field evolution in figures \ref{fig:3} and \ref{fig:4}. 

 Regarding the core evolution, \cite{Castillo2017} have found that due to ambipolar diffusion the toroidal field is driven inside closed poloidal field lines.  This behaviour is also observed in simulations where we consider a neutron star without a crust and a core occupying the entire star. In the simulations presented here however, due to the simultaneous evolution in both regions this effect is not completely evident, although a movement of the toroidal field  towards the regions of closed poloidal field lines can be seen in the first few kyr of the evolution. However as the evolution in the crust proceeds, the region of closed poloidal field lines is pushed completely to the crust where the mechanism of ambipolar diffusion is not operating. Thus, in the timescales studied here toroidal field is also present outside the region of closed poloidal field lines.

\begin{figure} 
    \includegraphics[width=.46\textwidth]{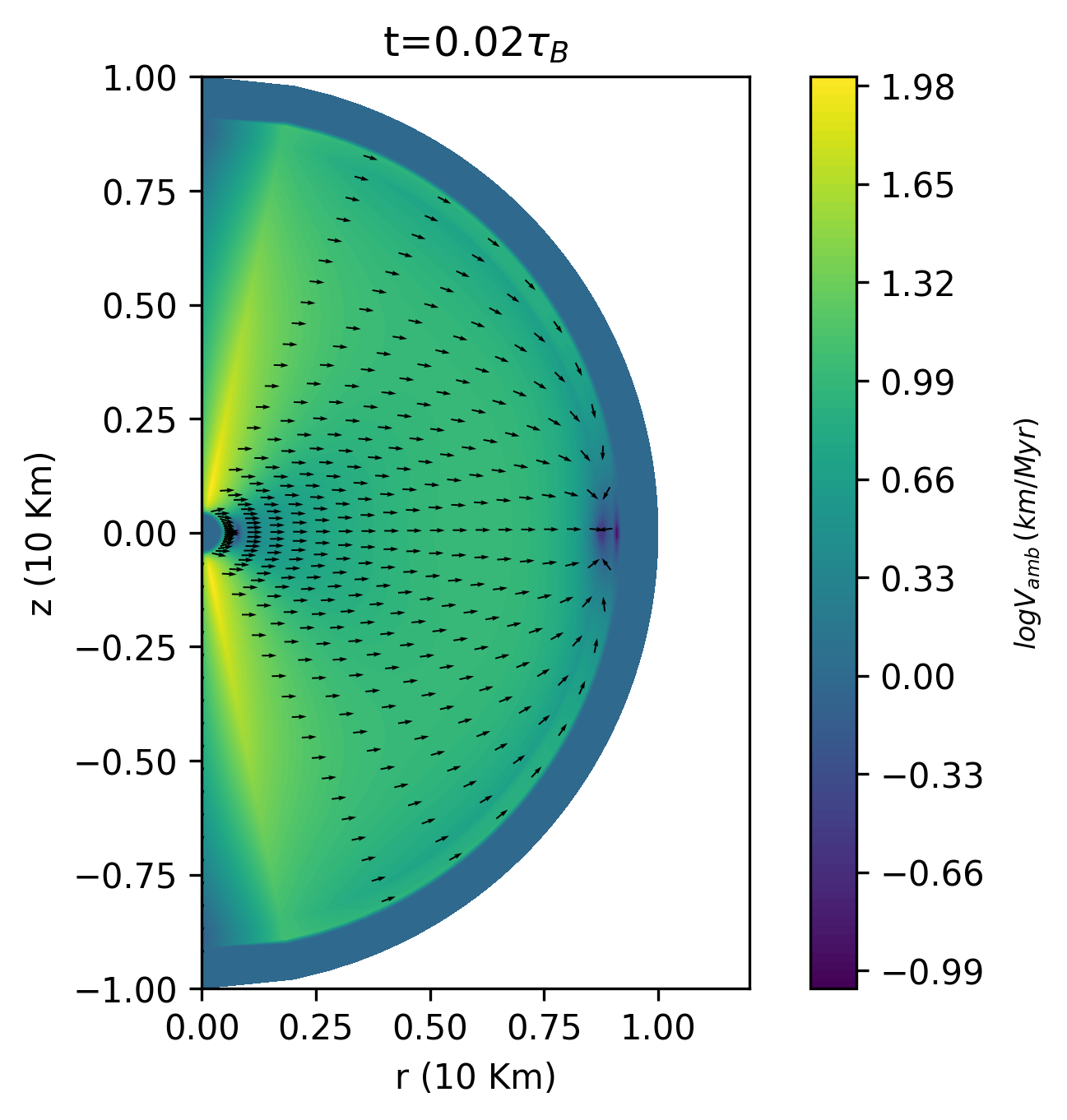}
    \caption {The velocity of ambipolar diffusion in the case of the purely poloidal initial magnetic field at the $t=0.2\tau_B$. The arrows indicate the direction of the velocity on a meridional plane while the color is the $\log|\bmath{u_{amb}}|$. }
    \label{fig:8}
\end{figure}

The values of $t_{pn}$ and $\lambda$ used here, correspond to a temperature of $10^9 K$. At this temperature the ambipolar velocity has a strong irrotational part as the relevant reactions are fast enough and so the magnetic field is advected to the exterior of the core towards the center of the closed poloidal field lines \citep{Passamonti}. That is the case in our simulations with the outward movement of the magnetic field with a velocity profile like the one shown in figure \ref{fig:8}.

The departures from chemical equilibrium in this range of temperatures are small enough because the reactions are sufficiently fast and the equilibrium is restored quickly. Thus, the irrotational part of the magnetic force will not be completely balanced by the gradient of the departures from the chemical equilibrium and the velocity of the ambipolar diffusion will have a large irrotational component. 

At lower temperatures the reactions will not be fast and the departures from chemical equilibrium will be larger developing a gradient that can suppress better the irrotational part of the magnetic force. At such temperatures the values of the physical parameters result to a smaller timescale for ambipolar diffusion making the evolution of the core faster and to the same order with the timescale of the Hall effect. In simulations with such temperatures the end state configuration is reached faster and the most of the toroidal field, that was initially in the core, is expelled to the crust-core interface. 

However, special care should be taken with the boundary condition for the departures from chemical equilibrium at the crust-core interface. Enforcing the velocity of ambipolar diffusion to be zero at the crust-core interface can result in large gradients of $\Delta\mu$ and the appearance of numerical artifacts on the magnetic field evolution. 

The Hall effect does not dissipate the energy of the magnetic field of a neutron star, thus the mechanisms that contribute to the energy dissipation are the Ohmic dissipation and ambipolar diffusion. The timescale of Ohmic dissipation (equation \ref{ot}) is independent of the strength of the magnetic field and scales with electrical conductivity. As a result in a core of a neutron star the timescale of the Ohmic dissipation would be very long to provide alone a path for the magnetic field energy dissipation in a reasonable timescale. Without considering the effects of superfluidity and superconductivity, ambipolar diffusion, through the frictional drag between the charged and neutral particles and neutrino and anti-neutrino emitted from the beta reactions, provides a viable path for the energy losses. 

Additionally the behaviour of ambipolar diffusion pushing the magnetic field towards the crust of the neutron star, a region with lower electrical conductivity than the core, can enhance the energy losses from the Ohmic dissipation  by transporting the magnetic energy there. In highly energetic neutron star with strong magnetic fields, like magnetars, ambipolar diffusion combined with the presence of strong toroidal fields should be considered as a viable deposit for efficient energy dissipation in the depths of the crust. 

Several works have performed studies of the coupled evolution of the crust and the core magnetic field,  with a variety different assumptions for the physical mechanisms operating, we will perform below a comparison between these studies and our results. 
In \cite{Bransgrove2017}, the evolution of the magnetic field was studied both in crust and the core of a neutron star provided that the core reaches a hydromagnetic equilibrium in response to the evolution in the crust. The core magnetic field structure is given through the implementation of hydromagnetic equilibrium; we note however that ambipolar diffusion is not accounted for. Under the assumptions employed in that work, the initial poloidal core magnetic field is parallel to the axis of symmetry, forming straight lines, and after sufficiently long evolution it accumulates close to the crust-interface, due to flux tube drift \citep{Jones2006}. In our approach, we find that the poloidal core magnetic field approaches the straight line and parallel to the axis structure due to ambipolar diffusion in a scale of $
\sim 500$ kyr. 
Furthermore, in our results the toroidal field tends to be nested within poloidal flux loops, whereas in \cite{Bransgrove2017} the toroidal field is present mainly in the crust, with some toroidal field in the core if relaxation to a hydromagnetic equilibrium is postulated. As our aim is to examine how the interplay between the ambipolar diffusion and the Hall effect will impact the global evolution, we not require the core to reach the hydromagnetic equilibrium. In the work of \cite{Elfritz2016} several effects are accounted for the flux-tubes in the core, including the buoyant force, superconductivity, the drag forces and their self-tension, but not ambipolar diffusion. At the end of their simulation, the changes of the core field were very moderate and no indication of straightening of the poloidal magnetic field lines was noted, as opposed to our work.

\cite{Vigano2021} performed magnetothermal simulations for the first 100 kyr of evolution. In their study, they did not observed a significant evolution due to ambipolar diffusion in the core for the case of the toroidal field confined in the close poloidal field lines. There was only a small bending of the poloidal field lines accompanied with a small displacement of the toroidal field outwards. Similar behaviour was observed in our simulation for $t<0.05\tau_B$ and the temperature of $10^9 K$. As the main aim of their work was on the magnetothermal evolution they did not explore a wide range of initial conditions, other than the twisted torus one, in their runs with ambipolar diffusion.

For lower temperatures the velocity of ambipolar diffusion is higher and the effects of ampipolar diffusion stronger. However as noted in \cite{Vigano2021} and we also encounter in our simulation numerical instabilities developed on those temperatures due to strong gradients around crust - core interface. In order to deal with this issue a more detailed approach is needed for this region and it will be addressed in future research.

\section{Conclusions}

The combined evolution of the magnetic field of neutron stars with Hall effect and Ohmic dissipation in the crust and ambipolar diffusion in the core has been examined in this work. 

Ambipolar diffusion in the core of the neutron stars pushes the magnetic field toward crust - core interface. Meanwhile, in the crust toroidal field is generated due to the Hall effect. Moreover, the presence of an internal toroidal field in the form of a twisted torus breaks the north-south symmetry. Ambipolar diffusion makes the poloidal field lines parallel to the star's axis and traps the toroidal field into closed loops in the core. As the field dissipates the poloidal field lines move outwards while the toroidal field south or north depending on the initial polarity. 

In all the cases studied here the evolution relaxes to a configuration strongly influenced by the evolution of the crust. In the cases where poloidal field exists the field relax to a state similar to the Hall attractor in the crust. A purely toroidal field relaxes to a state with the remaining field moving towards the north or south depending on the polarity of the initial field as determined by evolution due to the Hall effect.

For the general case a larger variety of physical parameters  can be examined to derive general conclusions. Furthermore, a significant role in the outcome has the modeling of the crust - core interface and the condition for the matter that will be applied there. A simplistic assumption for the transition of the evolution from core to crust could constraint the parameter space as well as the numerical capabilities and the accuracy of the code  by developing strong gradients.

Regarding energy dissipation, configurations with enough energy in their toroidal fields result in faster magnetic energy dissipation. This kind of magnetic field configurations might exist in magnetars where large energy releases required for their activities.

Although the results outlined here are constrained by the assumptions of constant temperature, piece-wise homogeneous physical parameters, immovable neutrons and axial symmetry, they reveal the basic characteristics of the evolution in each region have in the global magnetic field evolution. Furthermore, the choice of a constant temperature impacts the ambipolar velocity profile. Selecting a lower temperature results to a stronger irrotational component of the ambipolar velocity profile and thus an impact on the combined evolution. We note however that the impact of the temperature on the evolution is left for future investigation. At lower temperatures however and below a critical temperature the effects of superfluidity and superconductivity should also be considered.

Overall, the combined evolution of ambipolar diffusion with the Hall effect leads to efficient decay and conversion into thermal energy. The core field evolves towards a uniform field with straight field lines in the core which is then linked to the crust field. The crustal field generates a toroidal component. As the toroidal field moves towards the crust-core boundary, on either side of the interface, fields of opposite polarity create sites of strong dissipation due to the large currents that may trigger current-driven instabilities in this region.

\section*{Acknowledgements}
We thank A. Igoshev and A. Reisenegger for insightful comments on the manuscript. We would also like to thank the anonymous referee for useful comments that helped to improve the manuscript.
DS is financially supported by the "Andreas Mentzelopoulos Foundation". KNG acknowledges funding from grant FK 81641 "Theoretical and Computational Astrophysics", ELKE.

\section*{Data Availability}

The codes and the data that were used to prepare our models
within the paper are available from the corresponding authors upon
reasonable request.


\bibliographystyle{mnras}
\bibliography{example} 




\bsp	
\label{lastpage}
\end{document}